%% file: article.tex
\documentclass[]{spie}  

\input{globalcommands.tex}

\usepackage{amsmath,amsfonts,amssymb}
\usepackage{graphicx}
\usepackage{setspace}
\usepackage{tocloft}
\usepackage{xcolor}
\usepackage{lineno}
\usepackage[normalem]{ulem}
\title{The Far-Infrared Enhanced Survey Spectrometer (FIRESS) for PRIMA: Approach and Estimated Performance}

\author[a,*]{C.M. (Matt) Bradford}
\author[b]{Alan J. Kogut}
\author[b]{Dale Fixsen}
\author[a]{Klaus Pontoppidan}
\author[a]{C. Darren Dowell}
\author[b]{Jason Glenn}
\author[a]{Thomas Pagano}
\author[a]{\textcolor{black}{Joseph Green}}
\author[a]{Marc Foote}
\author[a]{James McGuire}
\author[f]{Michael Rodgers}
\author[a]{Robert Calvet}
\author[a]{Hien Nguyen}
\author[g]{\textcolor{black}{Steve Hailey-Dunsheath}}
\author[g]{\textcolor{black}{Logan Foote}}
\author[g]{\textcolor{black}{Elijah Kane}}
\author[a]{\textcolor{black}{Reinier M.J. Janssen}}
\author[a]{\textcolor{black}{Margaret Meixner}}
\author[c]{Alexandra Pope}
\author[d]{Alberto Bolatto}
\author[e]{JD Smith}

\affil[a]{Jet Propulsion Laboratory, California Institute of Technology, Pasadena, CA 91109}
\affil[b]{NASA Goddard Space Flight Center, Greenbelt, MD}
\affil[c]{University of Massachussets, Amherst, MA}
\affil[d]{University of Maryland, College Park, MD}
\affil[e]{University of Toledo, Toledo, OH}
\affil[f]{Synopsis Inc, Pasadena, CA}
\affil[g]{\textcolor{black}{California Institute of Technology, Pasadena, CA 91125}}

\cftpagenumbersoff{figure}
\cftpagenumbersoff{table} 
 \usepackage{graphicx,makecell}
\begin{document}

\maketitle

\begin{abstract}
We present the architectural concept for the Far-Infrared Enhanced Survey Spectrometer (FIRESS) for the Probe Mission for far-IR Astrophysics (PRIMA).  FIRESS spans the 24--235\mm\ range with four \Res~$\sim$100 slit-fed grating modules, each coupling to a 24 (spatial)$\times$84 (spectral) pixel array of kinetic inductance detectors (KIDs).  All four arrays are read out simultaneously, and a point source of interest can be coupled to two of the four bands at a time.  A Fourier transform module can be engaged over a portion of the FIRESS slits to create a high-resolution mode in which the light is intercepted, processed by the interferometer then reinserted into the path to the grating modules for detection.  We provide a simulation and description of the technique that will be used to obtain high-resolution spectra.  We identify the most important system requirements imposed by the detector system, finding that they are met with the existing design.  Finally, we present our performance  \textcolor{black}{modeling, including both direct estimates given our current design status, as well as durable guidelines for developing general-observer programs.}
\end{abstract}


{\noindent \footnotesize\textbf{*}C.M. Bradford:  \tt{bradford@caltech.edu} }


\section{Introduction}
\label{sect:intro}  
The far-IR waveband \textcolor{black}{($\sim$25 to 250~\mm)} is the repository of much of the energy released in galaxies across their history \textcolor{black}{\cite{MadauDickinson_14}}, \textcolor{black}{indicating that much of the energy production has been in regions obscured to UV and optical light.} Dust forms quickly in the Universe, \textcolor{black}{with several observations showing large reservoirs in the first billion years\cite{Watson_2015_earlydust,Tamura_2019_oiii_dust,Witstok:2023_carbon_dust}}.  \textcolor{black}{Looking across the arc of history,} dust-obscured star formation appears to \textcolor{black}{have been} the dominant mode during the epoch when the universe was most active \textcolor{black}{between redshifts 3 and 1 \cite{Zavala_2021_IRLumFun}. As such, dust-obscured star formation is the dominant contributor to the Universe's integrated stellar populations (co-moving star formation rate density $\times$ time).}  There is evidence that a similar situation may \textcolor{black}{have been} at play for black hole growth, with a large but uncertain fraction of the integrated history \textcolor{black}{likely to have been}  obscured by dust \textcolor{black}{\cite{Hickox_Alexander_2018}}.  

Spectroscopy at rest-frame mid- to far-IR wavelengths offers an excellent \textcolor{black}{probe of this dust-obscured component of evolving galaxy populations.}   Like the UV / optical / near-IR, the far-IR includes spectral features that \textcolor{black}{originate in the ionized media around young massive stars\cite{Stacey_93}} and accreting black holes \cite{Sturm_2002,Stone_2022}.  The ensemble of mid to far-IR transitions measures the instantaneous rate of growth of stellar populations and black holes, as well as elemental abundances resulting from integrated star formation and feedback processes.  Unlike the shorter-wavelength spectral tracers however, the mid- and far-IR features are \textcolor{black}{largely unaffected by} dust obscuration. 

\textcolor{black}{The dust-penetrating power of the far-IR waveband is also vital in the study of planetary system formation from disks.}  \textcolor{black}{Protoplanetary disks carry the ingredients that determine the ultimate composition of planets, including bulk gas, dust, water (vapor and ice), and well as other volatiles. These materials can be inventoried via their spectral features in the mid- and far-infrared.  In particular, the most basic property of protoplanetary disks--their mass--is not readily measured with millimeter-wave tracers accessible to ALMA, but is measured with the lowest two rotational transitions of hydrogen deuteride (HD)\cite{Bergin_13}.  HD is a chemical analog of \hh, but unlike \hh\, HD has a dipole moment, and with a first rotational level energy of 128~K, is excited in typical disks. The transition has only moderate optical depth and thus provides a measure of \hh\ mass, particularly when combined with other tracers\cite{Seo_2024_disksHD}.  This \jone\ transition of HD is at $\lambda$=112\mm, and is a key target for PRIMA spectroscopy.  Another key question is the mass and temperature distribution of water vapor.  Although JWST spectroscopy has revealed rich evidence of the water content in disks \cite{Banzatti_2025_JWSTdiskWater}, JWST mainly probes the warm water component; most of the water vapor is in cool components, which emit at wavelengths beyond the 28-micron cutoff of JWST MIRI.  Further detail on these and other spectroscopic studies with FIRESS on PRIMA can be found in the article by Pontoppidan et al. in this issue.}       

Recognizing the \textcolor{black}{broad promise of the far-IR}, Astro2020 
recommended that a far-IR (or X-ray) Probe be built for science operations starting early in the 2030 decade.  
Revolutionary advances are possible even at probe scale in the far-IR, because we have not yet fielded a platform that operates at the fundamental limits posed by the solar-system and Galactic dust in the waveband between JWST and ALMA.  While this has been articulated for some in the context of the SPICA\cite{Nakagawa_02,Bradford_06,Roelfsema_18} and Origins\cite{Meixner_19_Origins,Bradford_2021_OSS} studies, the potential advances may not be universally appreciated in the astronomical community.  To understand the opportunity at hand, consider that relative to the existing far-IR instruments, 
a cryogenic telescope in space offers a reduction in background that is analogous to that obtained at optical wavelengths when comparing the daytime sky to the moonless night sky at a good site.  Spectroscopy is a particular beneficiary of the low-background platform, if detector sensitivities are good enough to take advantage of it.  This is now the case, and 
\textcolor{black}{we have developed the PRobe Mission for far-Infrared Astrophysics (PRIMA) to realize these scientific opporutnities. PRIMA employs a 1.8-meter telescope cooled to 4.5~K designed for an halo orbit about the earth-sun 2nd Lagrange point.  PRIMA has two instruments: a hyperspectral and polarimetric imager called PRIMAger, and FIRESS. The mission as a whole is described further in an article by Glenn et al.~in this issue, and further information on the PRIMAger instrument and its scientific goals are presented in companion articles by Ciesla et al. and Burgarella et al.  In this article, we describe our design for FIRESS, the Far-Infrared Enhanced Survey Spectrometer. }

\text


\section{FIRESS Architecture and Observing Modes}

\subsection{Driving Principles}
FIRESS, and the entire PRIMA mission have been designed to balance the spectacular scientific opportunities available in the far-IR with the realities of a cost-capped mission, which we take seriously.  The key design priorities of FIRESS are 
\begin{enumerate}
\item Full-band coverage for all instrument modes, from 25\mm, where JWST MIRI performance is decreasing, to the onset of the atmospheric far-IR windows beyond $\sim$200\mm. 
\item Excellent point source sensitivity, which requires high efficiency and excellent detector sensitivity.
\item Large discovery capability, which translates to high mapping speed and ultimately to large array format.
\item Minimizing cost and risk.
\end{enumerate}

In terms of resolving power (\Res=$\lambda/\delta\lambda$), FIRESS emphasizes line detection capability, rather than measurement of line profiles for most applications.  This results in a much faster measurement, and thus the ability to survey more sources over the full spectral band to ultimately yield superior population statistics.  

\subsection{Top Level Approach}

At the heart of FIRESS are four slit-fed grating spectrometer modules that are logarithmically spaced in wavelength and combine to cover the full 24-235~\mm\ band.  Each provides resolving power between 85--150, and couples to an 84 (spectral) \by\ 24 (spatial) kinetic inductance detector (KID) array.  \textcolor{black}{Light entering the slit of each band is dispersed in the 84-pixel direction and detected across the full band simultaneusly.}  Further details are provided in Section~\ref{sec:gratings} below.  \textcolor{black}{All  pixels in all four bands---that is, the entire instrument---}are read out simultaneously.

\textcolor{black}{The FIRESS slits are aligned in pairs.  Bands 1 and 3 are aligned on the sky through the use of a dichroic filter.  Bands 2 and 4 are similarly aligned, but at a different field position than bands 1 and 3.  Thus a given point source of interest can be positioned to couple to the full spectral range of bands 1 and 3 simultaneously, or bands 2 and 4 simultaneously.  Figure~\ref{fig:FIRESS} (bottom, left) shows the slit configuration projected onto the sky.}
 This configuration allows for the use of two band-splitting dichroics with a generous transition region to enable high efficiency \textcolor{black}{for both the transmission and reflection bands in the dichroics}.  We considered using these dichroics in conjunction with a polarizing grid to allow all four bands to couple to the same source simultaneously, though with each band in single-polarization, as was baselined for the SPICA-SAFARI \cite{Jellema_17}and Origins-OSS \cite{Bradford_2021_OSS} spectrographs.  However, the demonstration of high-performance dual-polarization KIDs in the last few years \cite{HaileyDunsheath_18,Janssen2021JLTP,Hailey_2023,Day_2024} shifted the optimization in favor of the two-slit, dual-polarization architecture.  In addition to eliminating a costly optical element, the dual-pol approach offers both greater mapping speed (since all bands are more sensitive), and improved point-source sensitivity when only \textcolor{black}{two of the four} bands are required.  The penalty for full-band point-source measurements is very small: just the difference between the dual-pol vs optimized single-pol grating blaze efficiency, $\sim$10--25\%,  \textcolor{black}{and it is} mitigated by the higher efficiency and more fully background-limited detectors.

For high-resolution work, a Fourier-transform interferometer (called the Fourier-transform module, FTM) is engaged in the light path from the telescope to the grating modules.   When engaged with pair of mirrors on a single carriage, the FTM intercepts the incident light, processes it, and returns it to the grating modules for detection.  This basic approach was designed for both SPICA SAFARI\cite{Jellema_17} and Origins OSS \cite{Bradford_2021_OSS}, though for FIRESS we use a dual-polarization system. The FIRESS FTM offers \Res\ of up to 4,400 $\mathrm \times (112~\mu m/\lambda)$.  Further details on the approach, and notes on the trades we conducted considering other high-resolution approaches are presented in Section~\ref{sec:ftm}. 

\subsection{Observing Modes and Modulation\label{sec:modes}} 
\begin{table}[hb!]
\caption{FIRESS Modes and Modulation Frequencies.  PRIMAger also uses the scan map mode.} \label{tab:mod}
\begin{tabular}{|m{2.9cm}|m{5cm}|m{3.5cm}|m{4cm}|}
\hline 
Mode & Modulation Method & Required Mod. Freqs. & Available Mod. Freqs\\[0.2ex]
\hline\hline
Scan mapping\newline(low-res mode)& BSM scan and/or observatory scan, raster or Lissajous in 2D & \textcolor{black}{5}--100~Hz & 0.3--100~Hz \newline(PRIMAger: 0.3--175~Hz)\\[0.2ex]
\hline 
Low-res pointed\newline spectroscopy & BSM chop at \textcolor{black}{5} Hz along slit direction & \textcolor{black}{5} Hz + harmonics & Down to 1~Hz or below if needed\\[0.2ex]
\hline
High-res \newline spectroscopy (pointed) & \raggedright FTM scan \hspace{2in}(BSM and telescope fixed) & \raggedright 10--100 Hz \newline (\textcolor{black}{5} Hz $\times$ (235\mm/$\lambda)$ )& Down to 1~Hz or below if needed, higher frequencies possible in subarrays.\\[0.2ex]
\hline
\end{tabular}
Notes: \\
BSM = beam steering mirror\\
FTM = Fourier-transform module
\end{table}

\paragraph{Signal Frequencies} Like all far-IR / submm / mm-wave direct-detection instruments, FIRESS and its counterpart PRIMAger rely on modulation at audio band frequencies of $\sim$0.1 to 300~Hz for optimal sensitivity.  \textcolor{black}{Table~\ref{tab:mod} summarizes the modulation schemes and signal frequencies.} The low-frequency end of this range has been an important aspect of the experimental design particularly for mapping instruments.  It sets constraints on scanning for wide-field measurements, especially when surface brightness or intensity is of interest.  The low frequencies are typically limited by detector and electronics 1/f noise for space missions.  Excellent low-frequency stability has been achieved in the bolometers used in Herschel SPIRE and Planck HFI, and steady improvement is now being achieved in kinetic inductance detectors (see Day et al. 2024\cite{Day_2024} as well as \textcolor{black}{other recent articles} \cite{Hailey_2023,Foote_24,Kane_24}).  In our PRIMA pre-phase-A study, we adopted conservative minimum modulation frequencies for PRIMAger and FIRESS of 5~Hz and 10~Hz, respectively.   Given the recent progress with the KID low-frequency noise, we are now relaxing these to use lower frequencies in future design iterations.  Maximum frequencies will be limited by \textcolor{black}{sample rates, $f_{\rm max} = \frac{1}{2\,t_{\rm sample}}$, which for PRIMA is constrained by available downlink \textcolor{black}{, which allows for continuous 20 Mbps science data collection, incorporating} the bit depth required for each sample, and available compression algorithms. At present we baseline sample rates of 150--350~Hz for FIRESS low res mode, (double this in high-res mode since fewer pixels are read out), and 300--600~Hz for PRIMAger, but these are subject to further study of compression algorithms and required bit depths.}   \textcolor{black}{For reference, the detector time constant is} typically $\sim$1~ms (for the high-sensitivity FIRESS devices at low loading), corresponding to $\rm f_{\rm 3dB}\sim 160\, Hz$. 

\paragraph{Modulation Approach}
FIRESS and PRIMAger each use one of two identical 2-D cryogenic beam steering mirrors (BSMs) positioned at a pupil between the telescope and the slits. This system will be provided by the DLR and Max Planck Institute for Astronomy (O. Krause lead), and builds on the heritage of the cryogenic steering mirrors in Herschel PACS \cite{Krause_06} and SPIRE.  The BSMs can chop at 10 Hz with $>$80\% duty cycle \textcolor{black}{(fraction of time for which the boresight error is less than 1/10 of a pixel)} even at 25\mm, and execute 2-D scanning patterns for small maps.  \textcolor{black}{Further information on the BSMs can be found in the Glenn et al. PRIMA overview paper in this issue.  For FIRESS, the intersection of the telescope field of view and the BSM range is 20 arcmin by 13 arcmin, with the long dimension parallel to the slit length.}  For very large maps, the observatory  scan will be used, as with Herschel. The BSM can also be combined with observatory scan.   In the high-resolution mode, the scanning of the phase delay mechanism in the Fourier transform module (FTM) naturally creates modulation -- the signal at each wavelength is modulated with a sinusoid at its fringe rate, equal to v$_{\rm OPD}$/$\lambda$. Here v$_{\rm OPD}$ is the (constant) speed at which the optical path delay is changed, and can be varied as needed, for example to prevent undesirable audio frequencies from impacting a particular wavelength range (see Section~\ref{sec:ftm}).

\begin{table}
\caption{FIRESS Grating Module Specifications}\label{tab:gratings}
\begin{center}
\begin{tabular}{|l|c|c|c|c|}
\hline
Parameter & Band 1 & Band 2 & Band 3 & Band 4\\[0.2ex]
\hline\hline
Spectral range [\mm] & 24-43 & 42--76 & 74--134 & 130--235\\[0.2ex]
Spectral sampling [\mm] & 0.23 & 0.41 & 0.73 & 1.29 \\[0.2ex]
Resolving power (\Res=$\lambda/\delta\lambda$)& 95--150 & 85--120 & 90-125 & 95-130 \\[0.2ex]
Array format & \multicolumn{4}{c|}{24 spatial $\times$ 84 spectral, 900~\mm\ pitch}\\[0.2ex]
Pixel size on sky [arcsec] & \multicolumn{2}{c|}{7.6} & 12.7 & 22.9 \\[0.2ex]
Pixel pitch ratio to B1, B2 & \multicolumn{2}{c|}{ } & 5:3 & 3:1 \\[0.2ex]
\textcolor{black}{Slit width [arcsec]$^1$} & 10.2 & 11.4 & 19 & 34 \\
\textcolor{black}{Slit length [arcmin]$^2$} & 2.72 & 2.76 & 4.6 & 8.3  \\
\textcolor{black}{Slit separation [arcmin]} & \multicolumn{4}{|c|}{\textcolor{black}{3.3 arcmin between bands 2+4 and 1+3}}\\
\hline
    \end{tabular}
   
    \textcolor{black}{Notes: $1$: Slit width is an area of active optimization, values are subject to change, may be smaller. \\    
    $2$:  Slit length is the total useful length of both half-slits, not including the space between them.}
    \end{center}
\end{table}

\section{Base Grating Module Suite}\label{sec:gratings}

Table~\ref{tab:gratings} provides the important parameters for the four FIRESS grating modules.  All four are based on an identically-sized 24~$\times$~84 pixel focal plane, comprised of two 12~$\times$~84 pixel subarray chips with a 900-\mm\ pitch in a hexagonal-packed arrangement (see Section~\ref{sec:detectors}).  This makes the detector housings and suspensions common across all bands.   \textcolor{black}{While the pixel sizes in Table~\ref{tab:gratings} represents our current reference design from which sensitivities are computed, we are also considering a faster backend optical system which increases pixel sizes on the sky; this could offer improved point source sensitivity, particularly at the long-wavelength ends of bands 2,3,4 (see Figure~\ref{fig:pointed}), at the expense of some angular resolution.  Slit width can be optimized in Phase B and C with the benefit of stray light analyses and full diffraction simulations; the values here are likely oversized, which creates a conservative (meaning likely high) background noise estimate (see Section~\ref{sec:performance}).}

\begin{figure}[t!]
\begin{center}
\includegraphics*[clip,trim=65 155 25 135, height=8.0cm] {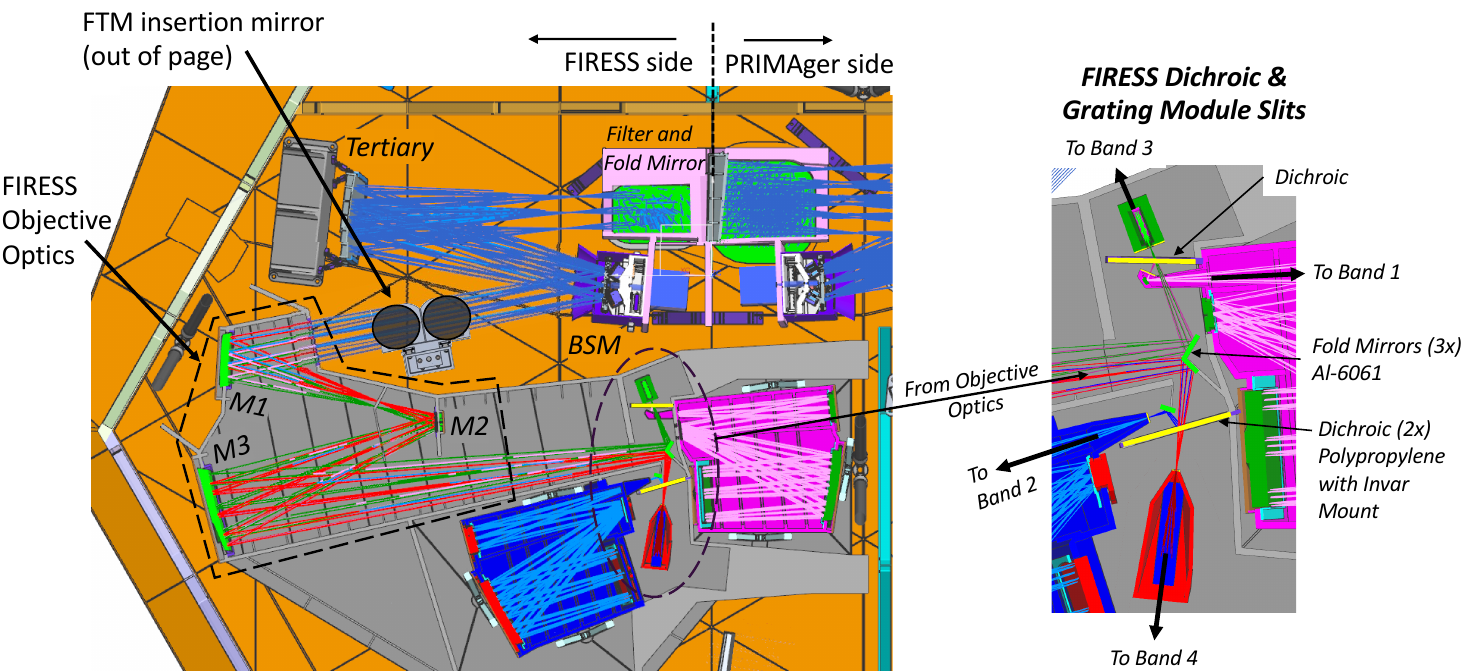}
\includegraphics*[clip, trim= 20 125 0 125, height=8cm]{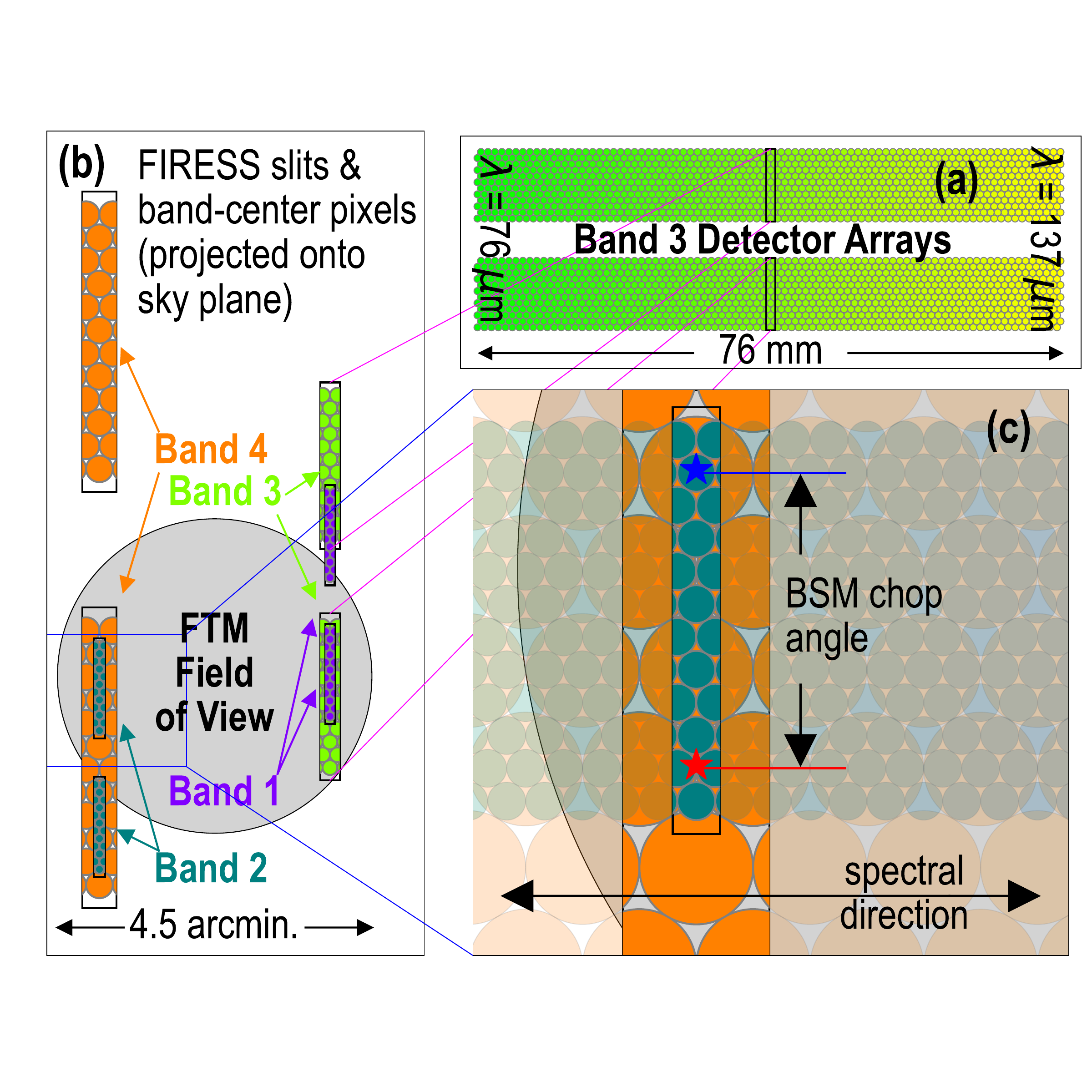}
\end{center}
\caption[FIRESS] 
   { \label{fig:FIRESS} 
\emph{Top:} FIRESS optical schematic, \textcolor{black}{looking toward the back of the telescope primary mirror. The telescope tertiary creates a collimated beam with an image of the primary at the beam steering mirror (BSM).  The path from the BSM to the FIRESS objective optics contains the FTM insertion mirror pair (see Figure~\ref{fig:FTM}).}  FIRESS' objective optics focus this collimated beam onto four slits for the four grating modules. Along the way, dichroic beamsplitters divide the 24--235\mm\ range into four bands (see Table~\ref{tab:gratings}). \emph{Bottom:} FIRESS field of view on the sky. Panel (a) shows the configuration of one of the arrays. All four arrays are 24 spatial (12 + 12) $\times$ 84 (spectral) pixels on 900-\mm\ hex-packed pitch.   Panel (b) shows the full FoV with all four slits. The slits are aligned to overlap in pairs, and the FTM field covers parts of all slits.  Panel (c) shows the alignment of pixels between bands 2 and 4, so that a chop along the slit can center a source in complementary spatial rows in both sides of the chop. }
   \end{figure}

\textcolor{black}{Moving spatially along the slit direction (vertically in Figure~\ref{fig:FIRESS} (bottom, right)), one encounters two flavors of spectral sampling, alternating back and forth.   Specifically, with a point source imaged to the position of the red star, the light is dispersed horizontally to the pixels to the left and right in the focal plane.  In wavelength space, the peak responses of the spectral pixels are determined by the x coordinate (spectral coordinate) of the center of these pixels.  The same is true if the source is moved up in the slit by 1 spatial pixel ($\sqrt{3}/2$ times the pixel spacing), except now the spectral coordinates of the pixel centers are shifted by 1/2 a spectral channel relative to those of the original position.   By positioning a source in both of these types of spectral rows sequentially, we can obtain improved spectral sampling relative to that of a rectangular array.   The black position shows an example of a row which is complementary to that of the red position.  This black-plus-red position pair represents an example chop pattern that would be used with the BSM.  }

Figure~\ref{fig:FIRESS} \textcolor{black}{(top)} shows the optical schematic on the left and the mapping of the slits and the pixelation of the array to the sky on the right.  The current design features a common plate scale for bands 1 and 2 which images at the band 2 diffraction limit. This is currently baselined for two reasons: 1) it allows the enclosure and most optical elements to be common between the two bands, which saves on design and manufacturing, and 2) the larger band-1 pixel provides some science resiliency in the face of non-ideal detector sensitivity and potentially in the system wavefront performance.   With the recent demonstrations of outstanding 25-micron KIDs \cite{Day_2024} since the step-1 proposal submission, this latter point has become largely moot, and this band 1/2 duplication is subject to a re-optimization in phase B.   Relative to the bands 1 and 2 pixel pitch, bands 3 and 4 have integer ratios of pixel pitches; this is so that a source can be centered on multiple spatial slit positions in both of bands 1 and 3 simultaneously, or bands 2 and 4 simultaneously, as \textcolor{black}{described above and} shown in Figure~\ref{fig:FIRESS} (right).  This 2-and-4 and 1-and-3 interband alignment is budgeted in our integration time estimates; an hour of integration includes 30 minutes in each of the two complementary spectral rows.

\begin{figure}[ht!]
\begin{center}
\includegraphics*[clip, trim= 650 130 320 240, height=7.3cm]{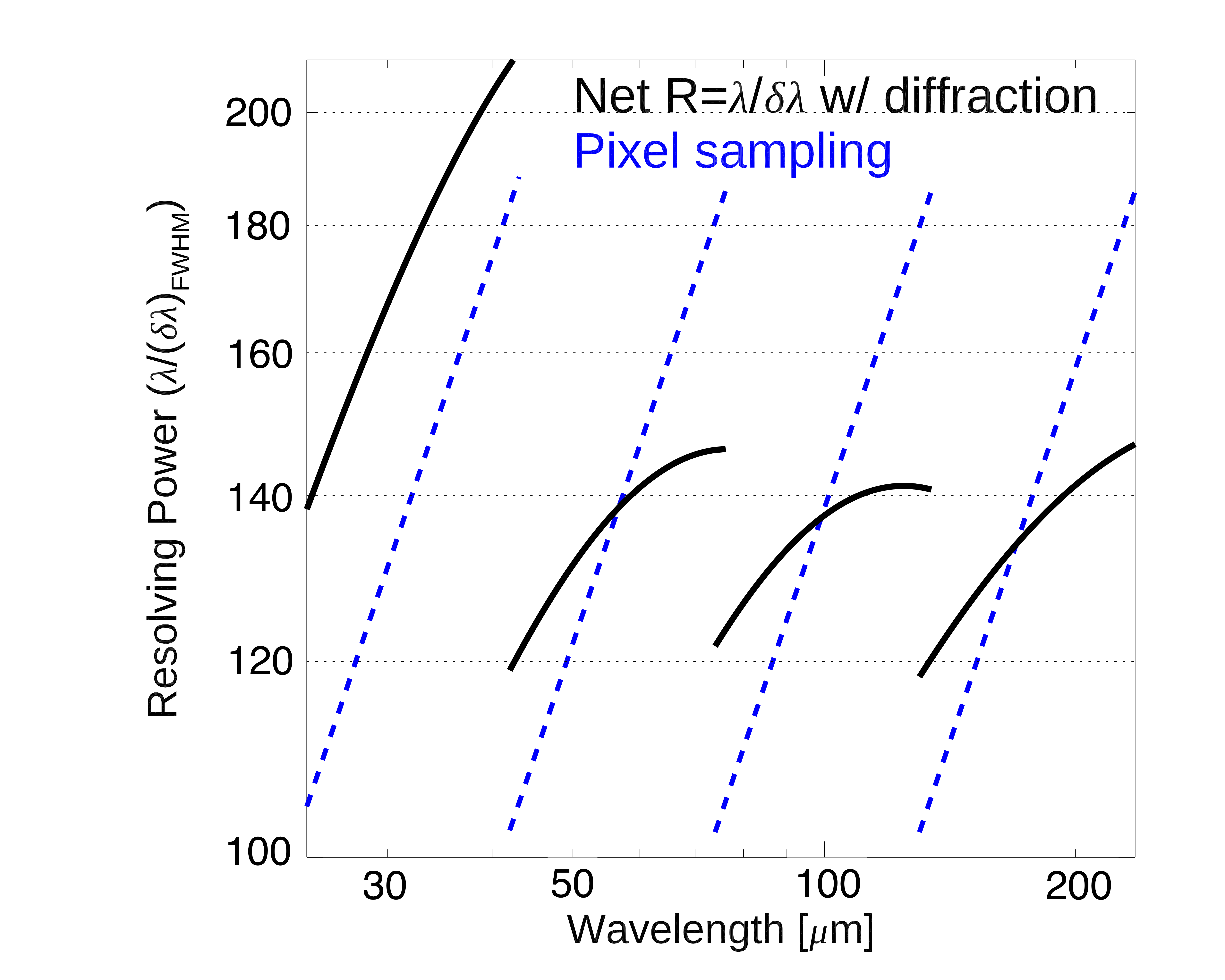}
\includegraphics*[clip, trim= 650 130 320 240, height=7.3cm]{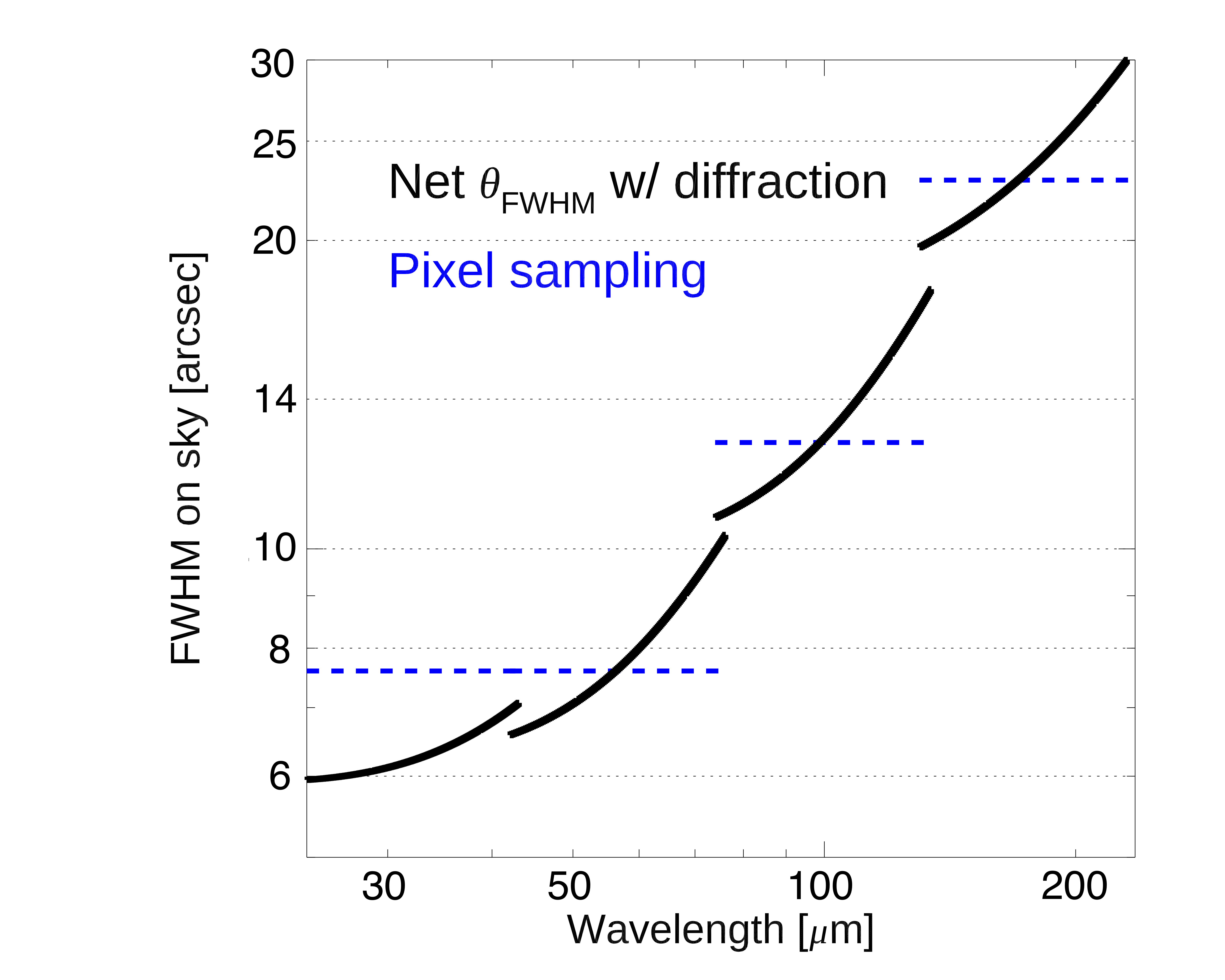}
    \end{center}
\caption[FIRESS] 
   { \label{fig:FIRESS_sample} 
FIRESS grating module resolving power \Res \textcolor{black}{(left) and on-sky beamsize (right)} across all four bands.  \textcolor{black}{All axes are logarithmic; vertical markers are 120, 140, 160, 180, and 200 at right and at 6, 8, 10, 14, 20, 25 at right)}.  The total resolving power is the convolution \textcolor{black}{of the pixel size and the diffraction in the spectrometer}.  Not shown but also included are beam broadening due to pointing errors and geometric aberrations, which are negligible.  Note that extended sources will have additional degradation due to the slit width, which is currently baselined to be \textcolor{black}{1.35 to 1.5} $\times$ larger than the pixel size.}
\end{figure}

Figure~\ref{fig:FIRESS_sample} shows the wavelength dependence of the resolving power (\Res=$\lambda/\delta\lambda$) \textcolor{black}{and on-sky beamsize} across all four FIRESS bands.  \textcolor{black}{These calculations rely on detail on the lens shape and KID absorber geometry (Section~\ref{sec:detectors}), and we have adopted the conservative position that the microlenses coupling light to the detectors have reduced response off center, dropping to zero at their edges. This assumption amounts to making the pixels smaller than their physical sizes (= their sampling).  This is conservative from a sensitivity standpoint (Section~\ref{sec:performance}), as it represents lower total point-source efficiency than we expect to obtain in practice.  However, it also creates higher \Res\ and smaller beams than may be obtained in practice, and is responsible for the spectral and angular resolving powers exceeding those set by the pixel sampling. Thus, we encourage caution with use of the parameters plotted in Figure~\ref{fig:FIRESS_sample} .  In any case,}  the half-pixel sampling  \textcolor{black}{provided by the hex-packed array architecture and the} chopping to complementary spectral rows ensures that the system can recover \textcolor{black}{the intrinsic \Res\ that a single pixel plus diffraction provides.}

The grating modules are made from aluminum optics inside an aluminum enclosure, with no moving parts, similar to the Spitzer Infrared Spectrograph \cite{Houck_04} and the JWST MIRI imager \cite{Bouchet_2015}.  Unlike the Spitzer IRS and MIRI, however, the FIRESS modules are cooled to 1~K with the upper stage of the PRIMA \textcolor{black}{continuous adiabatic demagnetization refrigerator (CADR)} (see DiPirro et al., this volume) to prevent thermal radiation from impacting the system sensitivity.  The focal planes (Section~\ref{sec:detectors}) and their bandpass filters are cooled to 120~mK with the lower stage of the CADR.

\section{High-Resolution Mode: Fourier-Transform Module}\label{sec:ftm}

For high-resolution spectroscopy, \textcolor{black}{the Fourier-transform module (FTM) intercepts the incident light, processes it, and returns it to the grating modules for detection. In this mode, PRIMA is pointed with the steering mirror fixed, delivering light from a source to a position on two spectrometer slits (Bands 1 + 3 or 2 + 4).}    As the optical path difference is varied, each grating spectrometer pixel measures its own interferogram, which is Fourier transformed to provide a subset of the spectrum.  The full spectrum is then created by stitching the sub-spectra together.  Two pointings per slit pair, to staggered spectral rows, provides a fully sampled spectrum.  This approach works identically to a broad-band Fourier transform instrument (e.g. COBE FIRAS \cite{Fixsen_94} or Herschel SPIRE \cite{Griffin_10}, except that the photon noise for any given part of the spectrum (which is what is relevant for line flux measurement) is reduced.

\subsection{Trade Study and Rationale}
We considered multiple approaches to high-resolution spectroscopy with FIRESS before arriving at the choice of the post-dispersed Fourier-transform architecture.  The most important scientific driver was the full-band instantaneous capability with good sensitivity, and sufficient resolving power for line detection.  At the same time, mass, volume, complexity, and thermal lift to support the 120~mK focal planes were key considerations.  Table~\ref{tab:highres} summarizes the study findings.  

\begin{table}[t]
\caption{FIRESS High-Res Trade Study}\label{tab:highres}
\begin{center}
\begin{tabular}{|p{3cm}|p{6.2cm}|p{6.2cm}|}
\hline
Approach & Favorable Attributes & Unfavorable Attributes \\[0.2ex]
\hline\hline \small
Fourier-transform (with grating post disperser) & \small Spectra obtained across entire range of multiple grating bands simultaneously.  Fastest method for full-band spectral survey to a given line sensitivity.  \Res\ selectable via scan length.  No additional detector arrays required.  \newline Herschel flight heritage.& \small Max \Res\ limited by physical scan: 4,400 $\times$ 112$\mathrm \mu m / \lambda$ for PRIMA FIRESS, thus can not kinematically resolve HD in disks.\\
\hline 
\small Scanning Fabry-Perot etalons (1 per grating module, grating is order sorter) & \small \Res\ up to $\sim$15,000 across PRIMA band with optimistic finesse (200).  Can tune across full band.  No additional detector arrays required. \newline ISO flight heritage. & \small Must scan to generate spectrum, resulting in sensitivity (or time) penalty.  Detector noise limits sensitivity, does not benefit from cold telescope.  At least 2 mechanisms per etalon (so 8 for all four PRIMA bands)\\
\hline \small
Virtual imaging phase arrays (VIPAs) with tunable cross disperser & \small High \Res\ is possible (up to 10$^5$ has been designed / proposed). No scanning required within small instantaneous band. & \small Instantaneous band limited by array size (e.g. 0.1\% at R=10$^5$).  Available wavelength coverage limited to $\sim$10\% with tuning mechanism in design.  \Res\ fixed in design, not selectable.  Detector noise limits sensitivity, does not benefit from cold telescope.  Requires additional detector array and large dedicated optics train for each band. \newline No far-IR flight heritage (but POEMM balloon experiment funded \cite{Nikola_VIPA_24}).\\
\hline \small
Heterodyne receiver & \small Very high \Res\ is possible (up to 10$^7$), and selectable.  & \small Large sensitivity penalty due to system temperatures much higher than cold telescope background in the far-IR.  Small instantaneous bandwidth (e.g. 3.5~GHz at 1.8~THz = 0.2\%).  Requires local oscillator injection / distribution in the focal plane.\\
\hline
    \end{tabular}
    \end{center}
\end{table}

\subsection{Implementation}
\begin{figure}
    \begin{center}
\includegraphics*[clip, trim= 200 150 200 150, height=8.7cm]{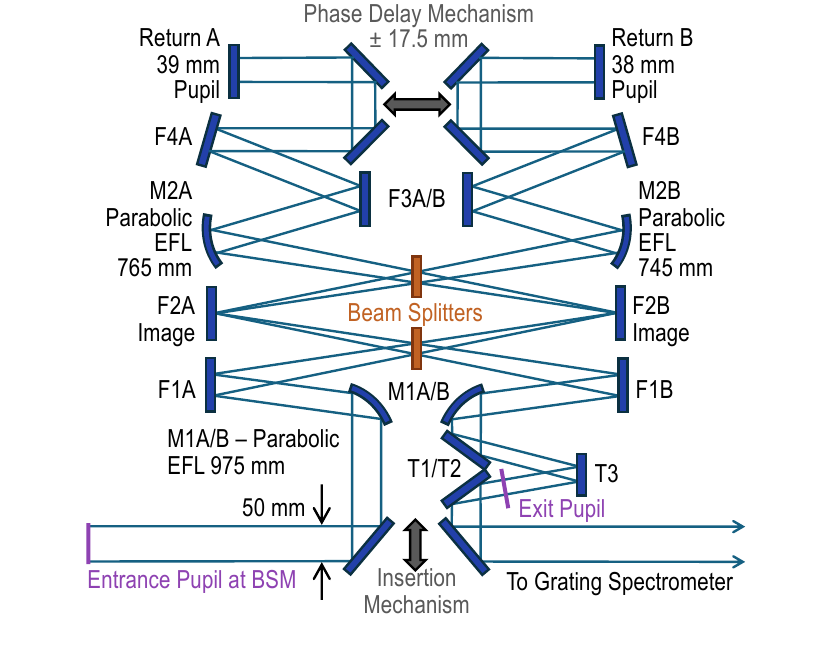},    \caption[FTM]{\label{fig:FTM} FIRESS Fourier-transform module schematic.  When the insertion mirror pair is engaged, the sky signal is intercepted on the way to the grating modules and processed by the FTM.  The use of only 2 ports, each serving as both input and output, and the two polarizing grids means that at zero path, the full dual-polarization sky signal is transmitted through to the gratings, and on average half the light is transmitted as the interferogram is obtained. Power is not shared with another output port.}
    \end{center}
\end{figure}

The FTM uses a linear mechanism to insert a pair of flat mirrors to divert the light into and out of the FTM.  The interferometer is a polarizing Martin-Puplett system \cite{Martin_70,Lambert_78}, with a pair of rooftop mirrors on a moving carriage to form the interferometer.   Figure~\ref{fig:FTM} shows the layout.  It is implemented in the manner of the FIRAS instrument on COBE\cite{mather_93} with two polarizing beam splitters oriented at 45 degrees to one another, and two ports (sky and cold grating modules) each serving as both input and output.  Aside from losses in mirrors and grids, the system delivers all the light in both polarizations from the sky to the grating modules at zero path difference (ZPD), and in general at arbitrary path difference, half the light is transmitted.  The moving arm is configured to create an 8$\times$ path multiplication from carriage motion to optical path difference (OPD) and to access the range from -20~mm to +260~mm OPD, providing a resolving power of 4,600 at 112\mm\ at full scan.  Lower resolving powers can be obtained with shorter scans.

The FTM has a 4.5-arcminute field of view, sufficient to provide \textcolor{black}{some} coverage in both slits, as shown in Figure~\ref{fig:FIRESS} (bottom, left).   Maintaining the full available resolving power across this FoV drives the pupil diameter according to:
\begin{equation}
    D_\mathrm{pupil} > \sqrt{R_\mathrm{max}\,\lambda\, D_{\mathrm tel}\,  \theta_{\mathrm sky}}
\end{equation}
as is true for any interferometer providing diffraction-limited imaging.  Here $\theta_{\mathrm sky}$ is the angle of the field edge from the boresight, 2.25 arcminutes.  Setting $R_\mathrm{max} = 9000$ at 112~\mm\ to insure divergence does not limit the performance, we arrive at a D$_{\mathrm pupil}>$3.5~cm.  This is comparable to the 5-cm pupil used in the telescope reimaging, and enables a modest package.  Powered mirrors are used in the interferometer to minimize beam walk, and the optical model shows a pupil overlap of 90\% at the most extreme OPDs (near zero and 260~mm), and better as the pupil match position of OPD=120~mm is approached. 

An interferogram is obtained by scanning the FTM optical path difference (OPD) at a constant speed from zero (white light position) to OPD$_{\rm max}$ and sampling the detectors at an interval $\delta t_{\rm sample}$ which corresponds to a OPD sample of $\delta x_{\rm sample}$.  Fourier transforming the interferogram creates a spectrum with synthesized channels in optical frequency running from 0 to $\nu_{\rm max}$ spaced by $\delta \nu$, where \begin{equation}
\nu_{\rm max}\, \left[\rm cm^{-1}\right]= \frac {1} {\delta x_{\rm sample}}, \hspace{2in} \delta \nu = \frac{1}{\rm OPD_{\rm max}}.
\end{equation}
As the interferogram is obtained, a given optical frequency $\nu$ maps linearly to an audio frequency $f$ according to \begin{equation}
f = \frac{\delta x_{\rm sample}}{\delta t_{\rm sample}} \times \nu = v_\mathrm{OPD} \times \nu
\end{equation}


With signal bands between 1 and 200 Hz corresponding to good sensitivity in our kinetic inductance detectors (KIDs) (Section~\ref{sec:detectors}), the target OPD scan rate is between 0.24 to 4.8 mm / sec in order to put fringe rates into the signal band for the full 25 to 235~\mm\ band. The ability to select the scan rate within this range provides some flexibility to avoid any bad audio frequencies for spectral lines of interest and overcome audio-band systematics if these issues exist.  For example, consider a Band~3 grating module pixel at 112~\mm.  In the current design, it has a spectral width of one part in 157 due to pixel sampling, and diffraction broadens the width to one part in 90.  The optical bandwidth of this pixel is thus between 2.662 and 2.691 THz.  For an OPD scan rate of 1 mm/sec, this narrow optical bandwidth translates to audio frequencies spanning only 8.87 to 8.97 Hz.  For this particular grating pixel, all signals from the sky are encoded within this narrow audio band.   The corresponding FIRESS grating pixel at 30 microns will have its signal encoded at frequencies between 33.1 to 33.5 Hz for the same OPD scan rate.  The longest-wavelength grating pixels in FIRESS, at 235\mm, have sky signals encoded at 4.23 to 4.28 Hz.  The narrow band content makes the system very robust to systematics and detector noise, especially when considering that $v_\mathrm{OPD}$ can be modified if desired for a particular observation.

PRIMA's FTM is designed to work with a demonstrated cryogenic double-flexure scan mechanism with the required 35~cm travel (\textcolor{black}{$8\times280\,\rm mm$)}, described in articles by Cournoyer et al. \cite{Cournoyer_22,Cournoyer_20}.  It exceeds the mechanical requirements for accuracy, stiffness, and scan velocity control.  The use of a stiffness-compensation scheme results in low power dissipation (0.4~mW max at 4~K), and low exported accelerations to the optical bench.

\subsection{Simulated FIRESS-FTM measurement}
\begin{figure}[t!]
    \begin{center}
\includegraphics*[clip, trim = 130 25 115 40, width=5.7cm,height=5.5cm]{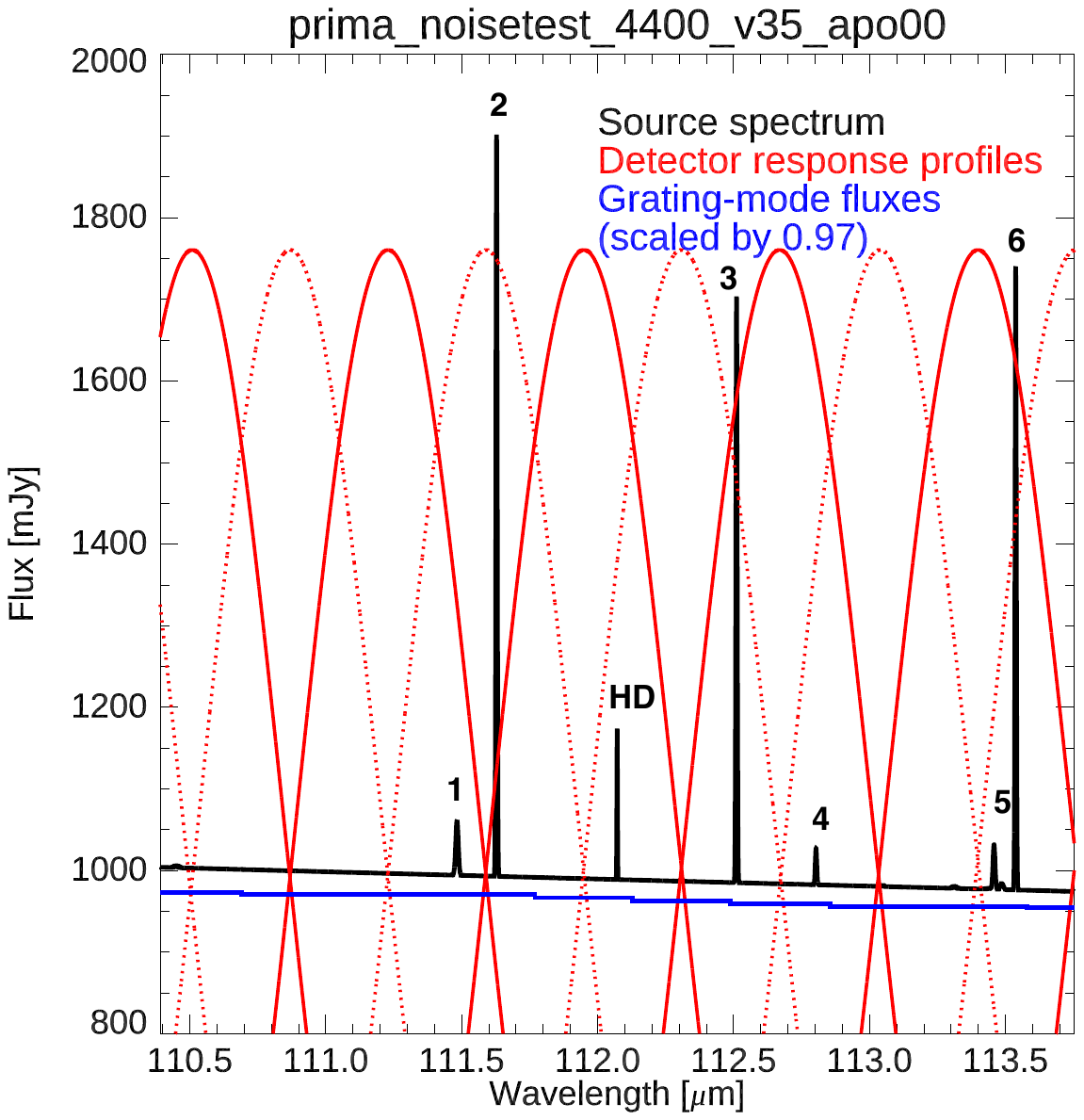}
\includegraphics*[clip, trim = 15 11 25 21, width=5.7cm,height=5.5cm]
{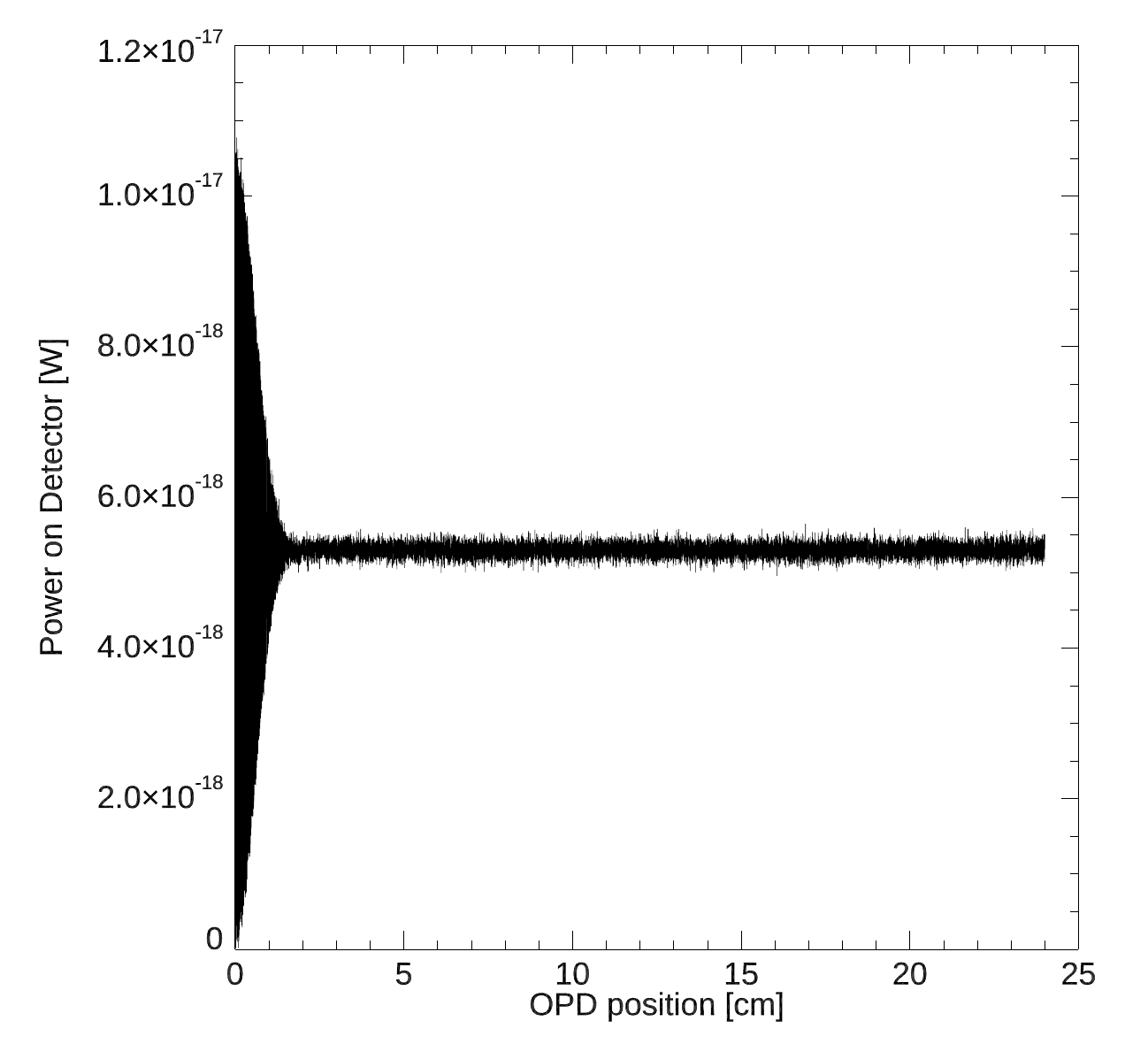}
\includegraphics*[clip, trim = 120 11 25 21, width=5.5cm,height=5.5cm]{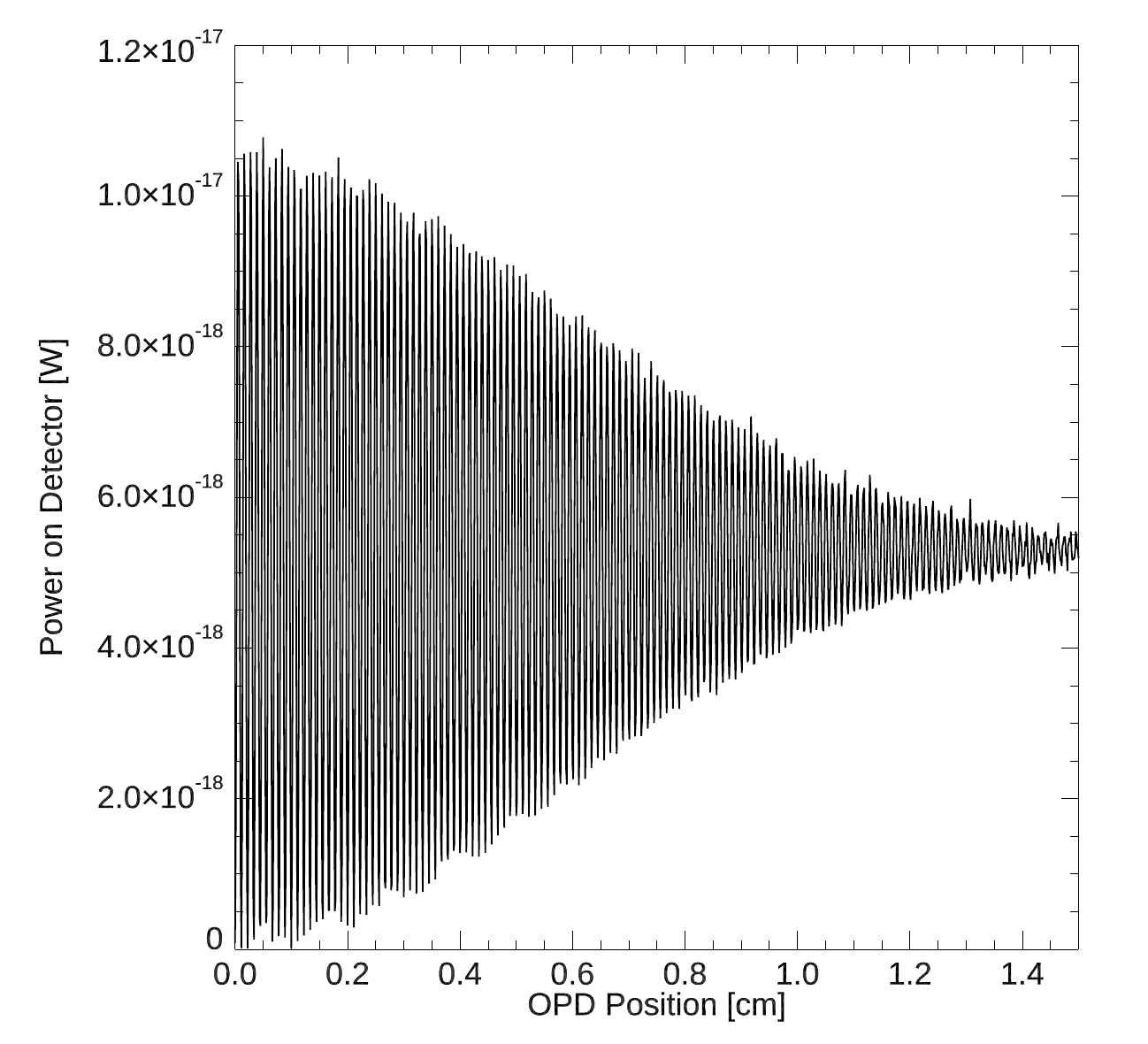}\\
\includegraphics*[clip, trim = 50 11 25 21, width=5.6cm,height=5.5cm]{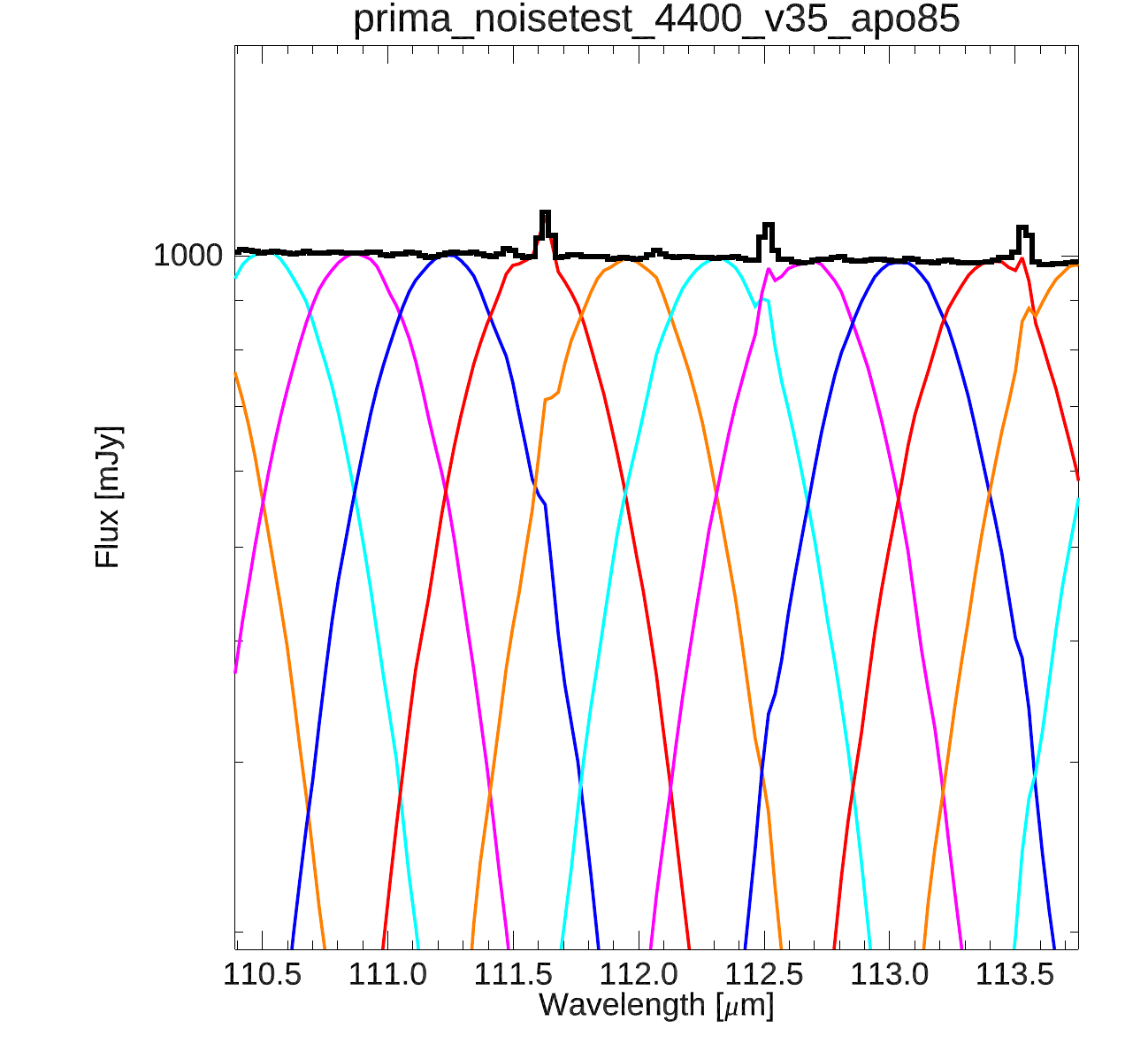}
\includegraphics*[clip, trim = 130 25 115 40, width=5.6cm,height=5.5cm]{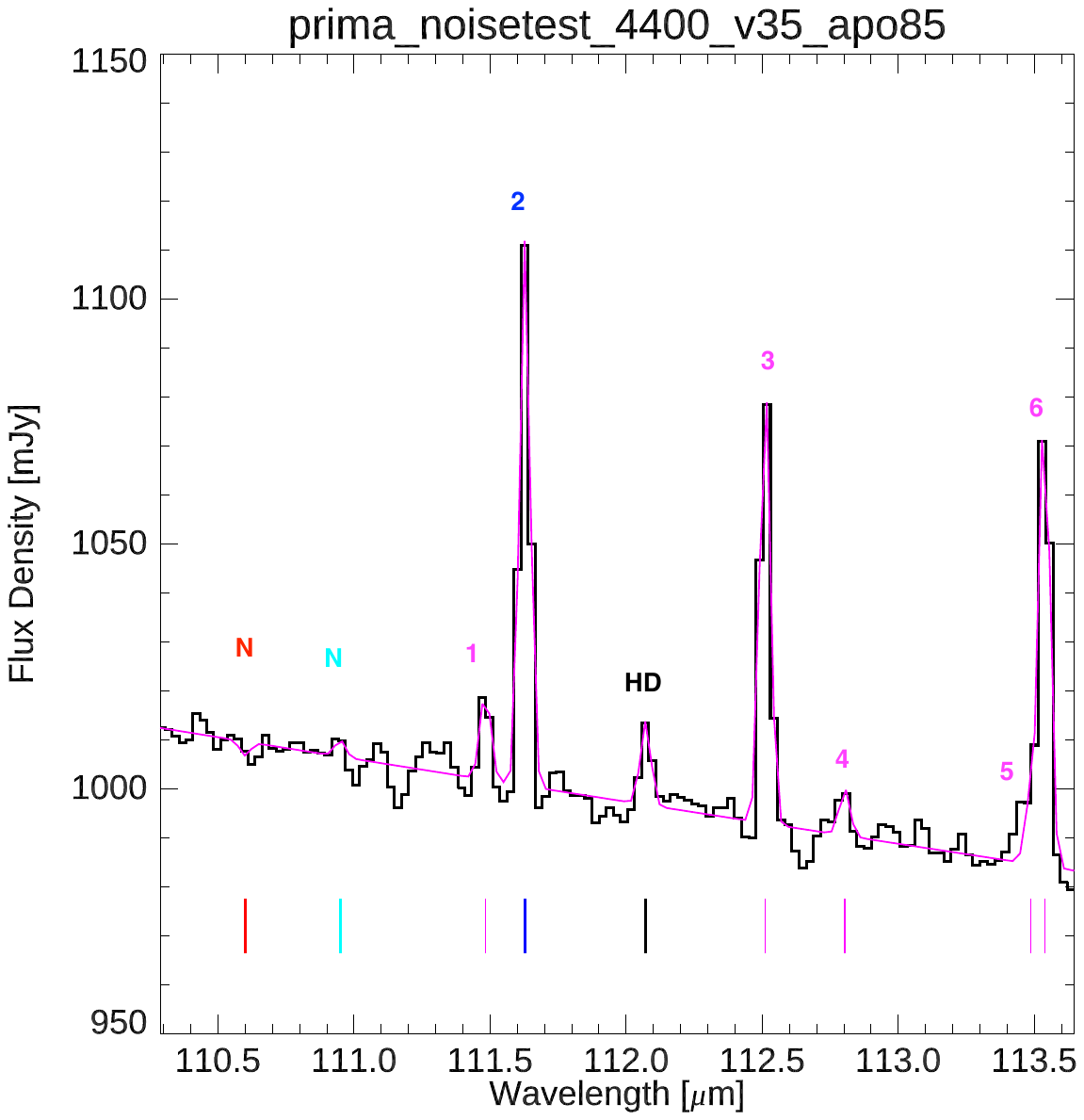}
\includegraphics*[clip, trim = 70 11 35 21, width=5.6cm,height=5.5cm]{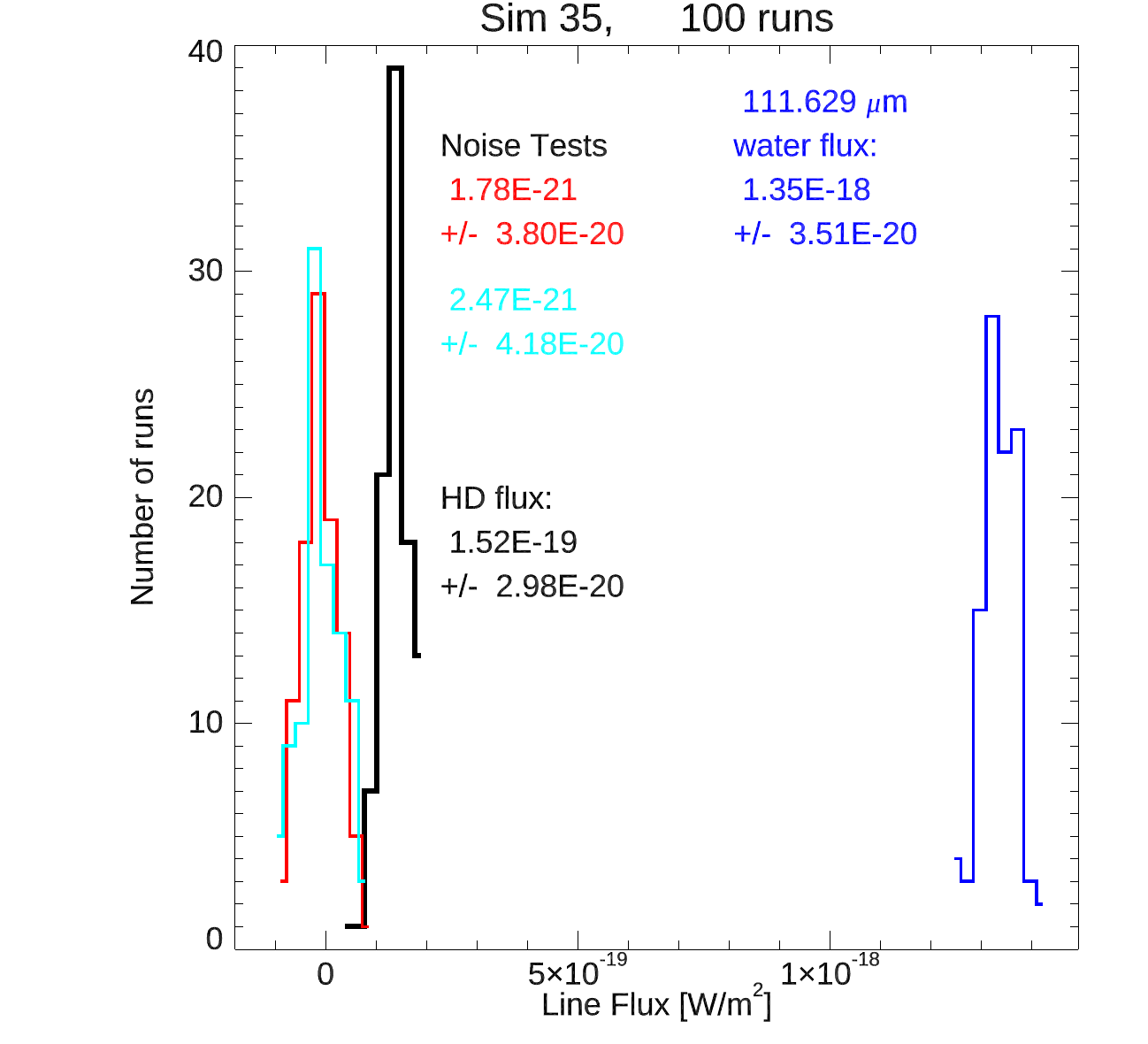}
    \caption[inteferogram]{\label{fig:interferogram} FIRESS FTM mock observation.  Top left shows a portion of a model spectrum in black with six water transitions (labeled 1-6) and the HD \jone\small\ transition at 112.07\mm.   Solid red shows the grating channel response functions for one spectral row, and black shows the grating channel fluxes.  The response function for the complementary spectral row with staggered spectral coverage is shown as dashed curves, these are not used in this simulation. Individual grating channels each record an independent interferogram as a function of the OPD scan position; an example is plotted on two scales in upper middle and upper right.   Lower left show the individual channel spectra (colored) obtained by Fourier transform of the interferograms.  The full spectrum in black above is obtained with an optimal weighting of the grating channels.  Bottom center provides a close look at the final composite spectrum and shows in magenta a fit to the power-law continuum and 10 line fluxes (identified with wavelength markers).  Here line wavelengths are held fixed and a common Gaussian width is fit.  The two shortest-wavelength `lines' with red and cyan N markers are dummy positions where no line flux is modeled; they provide an estimate of the uncertainties.  Right presents a histogram obtained by performing this mock observation and flux extraction 100 times: the red and cyan show the uncertainties estimated via flux extraction of the dummy lines which have zero input flux, while black is the HD \jone\small\ flux distribution, and black shows the bright water transition (transition 2).}
    \end{center} 
\end{figure}

The FIRESS FTM with the grating backends creates a post-dispersed Fourier transform system, an approach that is not widely appreciated in the general astronomical community (but was proposed for the SPICA SAFARI and Origins OSS instruments \cite{Gratton_92,Jellema_16,Naylor_22}).  We have developed a simulator for PRIMA FIRESS to demonstrate the operation and verify the sensitivity estimates used in the performance model, particularly for the case of faint lines against a bright continuum, as is needed for the HD measurement in protoplanetary disks.  Figure~\ref{fig:interferogram} shows this mock data flow for a small portion of the FIRESS band centered around 112\mm, \textcolor{black}{with noise based on a 10-hour observation}.  We begin with a subset of a model disk spectrum that represents a stressing case: 0.5 Jupiter masses of gas around a 1~\ms\ star at a distance of 140~pc.  The source signal is diffracted to a row of spectral channels spaced at one part in 160 as per the current band~3 design at 112~\mm.  The complementary row of spectral channels \textcolor{black}{are shown with dashed profile curves.  These are observed with the source pointed to a complementary spatial position within the slit (see Section~\ref{sec:gratings}, Figure~\ref{fig:FIRESS}).  The integration time in the simulation includes the time for both positions.} 

As the optical path difference (OPD) is scanned, each detector's interferogram is computed independently, with noise estimated dynamically at each interferogram position by combining in quadrature the intrinsic detector noise with the photon noise.  Figure~\ref{fig:interferogram} shows an example interferogram. Photon shot noise is the dominant noise term for most interferometer positions, it arises from the combination of background and source flux, and varies between zero and the full unattenuated value at the white-light position. (We employ a total efficiency from the telescope to the detector of 0.043 at the white-light position, consistent with the current best estimate (CBE) instrument model, and a detector intrinsic noise of 1$\times$10$^{-19}\,\rm W\,Hz^{-1/2}$.)  This simulation uses a total time of 8~hours, which would be generated via the co-addition of 11 to 110 individual interferograms, depending on the adopted scan rate.  The interferogram has high SNR around the white-light position, which shows $\sim$160 coherent oscillations consistent with the large continuum flux filtered through the grating channel bandpass.  At higher OPD positions, the high-resolution information is encoded, even though it is not visible in the raw interferogram.

These interferograms produce individual channel spectra on each detector in the spectral row, and the full spectrum results from combining the channel spectra with an optimal sum where each detector's signal is weighted as the square of its profile at each wavelength.  The spectral response function of the grating pixels must be well measured to complete this stitching, and this is an important part of our ground calibration plan.  It can also be measured in flight with the FTS itself, using bright spectrally-smooth sources such as asteroids. As with all FTS spectra, the detector noise manifests as fluctuations and ripples on scales corresponding to the spectral resolution.  Nevertheless, weak lines can be extracted from the spectrum, particularly when the wavelengths are known a priori, as is the case for the HD transitions in disks.  We run 100 such mock observations and fit line fluxes to provide an estimate of the uncertainties.  \textcolor{black}{The simulated spectra shown here include apodization of the interferograms, this is a choice that can be made in the reduction, after the observation.  In this simulation, we use a Norton-Beer function\cite{Norton_76} with an index of 1.5.} 

One subtlety to note in the simulation output is variation in the final sensitivity as a function of wavelength as one moves through the grating channels for a given pointing.  The best sensitivity is obtained at positions centered on one of the grating channels, since virtually all the signal \textcolor{black}{for the narrow spectral line} comes from a single detector in the final stitching; this is the case for the HD 112~\mm\ transition in this simulation.  In contrast, for positions between two grating channels, two detectors contribute approximately equally in the weighted sum performed in the stitching, and less total signal is captured; this is the case for the bright water line (\#2) in this simulation.  Here again, the staggered spectral rows will be used to build a stitched spectrum with uniform sensitivity, and the CBE performance estimates include a factor to account for the average effect.  The $1\times10^{-19}\,\rm W\,m^{-2}$ 1$\sigma$ sensitivity is consistent with the CBE estimate (see Section~\ref{sec:performance}), when including the \textcolor{black}{1}~Jy continuum, and scaling to the 8-hour integration time.

\section{Detector System }\label{sec:detectors}
\begin{figure}[t!]
    \begin{center}
\includegraphics*[clip, height=6cm]{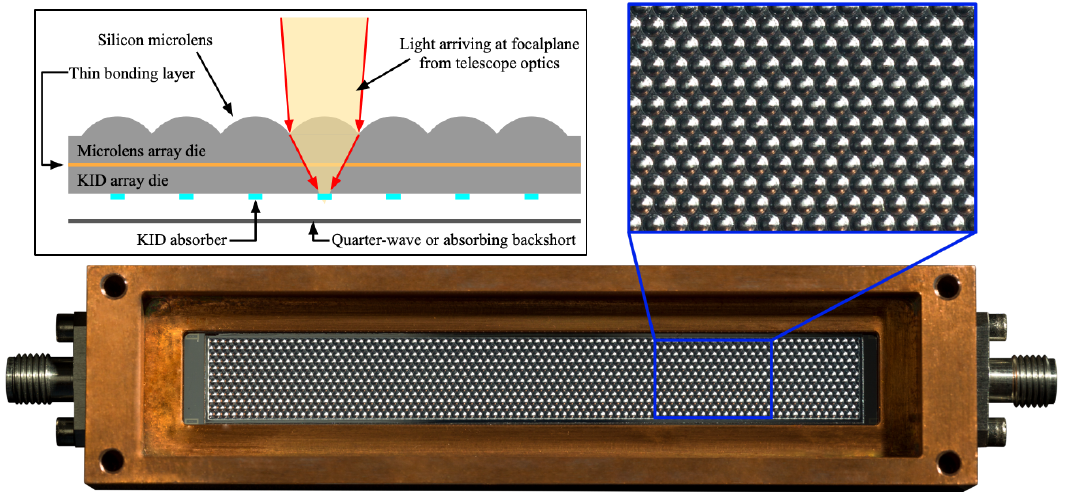}\\
\includegraphics*[clip, trim=0 2 0 0,height=5.cm]{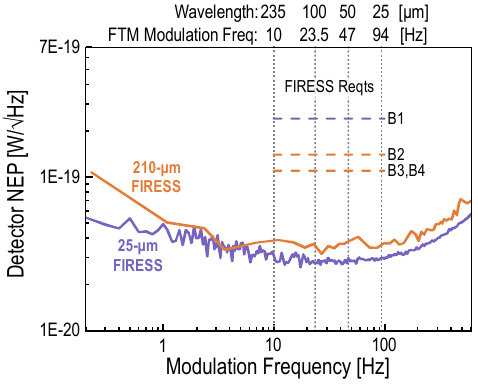}
\includegraphics*[clip, trim=0 2 0 0,height=5.cm]{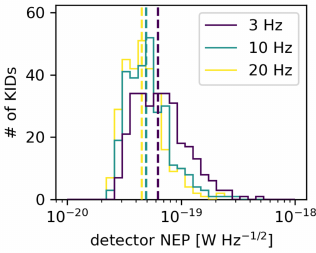}
    \caption[Subarray]{\label{fig:array}
FIRESS subarray architecture and performance.  
Top shows a single sub-array, lens side up, in a test housing. Bottom,left shows sensitivity measured as a function of modulation frequency for typical pixels in two FIRESS test arrays.  The requirements are set by photon noise in the four bands, but the required modulation frequencies were set conservatively; FIRESS will likely use frequencies down to 1~Hz and below to increase observing flexibility.  Bottom right shows a histograms of one recently-characterized die.  Of the KIDs available to our current readout system and fitting software, 92\% have NEP below 1$\times$10$^{-19}\,\rm W\,Hz^{-1/2}$ at 10~Hz. }
    \end{center}
\end{figure}

FIRESS uses aluminum kinetic inductance detectors (KIDs) which have have matured substantially in the last 2--3 years.  Here we provide a brief overview of the KID system design with an emphasis on how the arrays integrate with the instrument and impact the performance model.  Figure \ref{fig:array} shows the FIRESS focal plane architecture, one of the prototypes built to date, and recent sensitivity measurements.   Further details on the the most recent KID array performance, including the array-level measurements can be found in our peer-reviewed articles\cite{Hailey_2023,Foote_24,Day_2024}, as well as the recent proceedings now in press with the SPIE from the far-IR / submm/ mm-wave detectors portion of the Astronomical Telescope and Instrumentation conference.\footnote{Papers led by L. Foote, E. Kane, and C. Albert of Caltech / JPL discuss array measurements including matching resonant frequencies to pixel positions, and array sensitivity statistics.}  

Each of the eight 1008-pixel subarrays for the four FIRESS focal planes consists of a detector chip bonded to a matching microlens chip.   Details of the lens approach and bonding are described by Cothard et al.~2024 \cite{Cothard_24_lenses}.  The four bands will have different designs in the meandered inductors as well as the lenses, and we have thus far prototyped both KID arrays and lens arrays for 25 and 210 microns, approximately spanning the FIRESS range.  As Figure~\ref{fig:array} shows, both prototypes are showing excellent sensitivity which exceeds our requirements.  We are currently improving the yield in the 1008-pixel subarrays, and can report 95\% resonator yield in long-wave chips thus far.  Integration with readout is underway now -- we require 80\% yield to meet our science requirements.  

\subsection{KID Resonant Frequency Schedule and Readout}
For each 1008-pixel subarray, the KID resonant frequencies lie between 400~MHz and 2.4~GHz, spaced logarithmically for maximum separation.  With this schedule, the average fractional separation between adjacent KIDs in readout frequency is $\Delta f / f = 1.8\times10^{-3}$, as compared with a \textcolor{black}{maximum} expected fractional frequency shift under load of $\delta f / f <1\times10^{-4}$ \textcolor{black}{(For more information, see articles on the FIRESS KID modeling and testing \cite{Hailey_2023,Kane_24}).  Thus} we do not expect many dynamic crossings in the frequency ordering of the array as bright sources (and/or bright spectra lines) are moved through the array.  Each readout circuit consists of a mixed-signal electronics board in the spacecraft bus which generates a waveform with a tone for each KID, and receives the returning waveform after it interacts with the array.  The returning waveform is digitized and each tone's complex transmission (amplitude and phase) is extracted, revealing the frequency shift and the corresponding KID optical power.  The waveform is carried on a coaxial cable, which has a transition to superconducting cladding below 10 Kelvin.  A low-noise amplifier is staged at the observatory 18~K stage in the outbound run.  PRIMA has eight readout circuits to address the two subarrays in each of the four FIRESS focal planes.  All operate simultaneously, and the same readout electronics are used for the PRIMAGer instrument when it is operated.  PRIMA's readout electronics are led by GSFC and is described in greater detail by Essinger-Hileman et al.~ in an upcoming article.

\paragraph{Frame Rate} Among the key operational requirements for the readout is the frame rate (rate at which all the KIDs are sampled).  We set this at \textcolor{black}{9.5~kHz, a convenient value for the readout electronics, and} $\sim10\times$ the detector time constant to allow high-fidelity identification and removal of cosmic ray pulses on board prior to downsampling to the \textcolor{black}{downlinked science sample} rate of $\sim$ 150--600 samples per second \textcolor{black}{(see Section~\ref{sec:modes}).}

\subsection{Enclosure, Thermal Isolation, and Magnetic Shielding}
Both FIRESS and PRIMAger focal planes are operated at 120~mK in the baseline cryogenic system, which carries 100\% margin at all cryogenic stages\cite{Chen_23_PRIMAcryo}. With the aluminum film transition temperature of 1.3--1.5~K, cooling to 130~mK ($\sim\rm T_{\rm C}/10$), is sufficient to eliminate all thermally-generated quasiparticles (see Hailey-Dunsheath et al \cite{Hailey_2023}).  This both eliminates one source of detector noise, and  greatly reduces spurious signals created by temperature fluctuations; this is a key advantage of KIDs relative to bolometer systems.  Carrying the 120~mK design point thus represents a form of additional margin in the sub-K thermal architecture, as higher temperatures are readily accessible with easier system requirements.  The operating temperature will be optimized subject to the details of the as-built film T$_{\rm C}$ and two-level-system noise (TLS) in the devices, which is typically reduced at higher temperatures in the 100-200~mK regime.  

Temperature fluctuations are not expected to be a major problem.  There is a small TLS term which creates a temperature dependence of the KID's resonant frequency; its amplitude depends on the device but is of order $\rm (df/f) / dT \sim3\times$10$^{-11}$ per micro-Kelvin (see Kane et al. 2024\cite{Kane_24}, Figure~5).  With a modeled thermal noise on the focal planes of $<10^4\,\rm \mu K^2 \, Hz^{-1}$ (DiPirro et al., this volume), the corresponding fluctuations in fractional frequency (df/f) will be below 10$^{-17}\,\rm Hz^{-1}$. This is sub-dominant to the typical detector noise variance of $\sim3\times10^{-17}\,\rm Hz^{-1}$.  Further mitigation is provided by the fact that most of this temperature `noise' comes from short-period switching transients, with narrow-band frequency content; they can be easily identified and masked, and/or removed as a common signature across all arrays.

\begin{figure}[t!]
    \begin{center}
\includegraphics*[clip, trim=10 10 10 10,height=6.5cm]{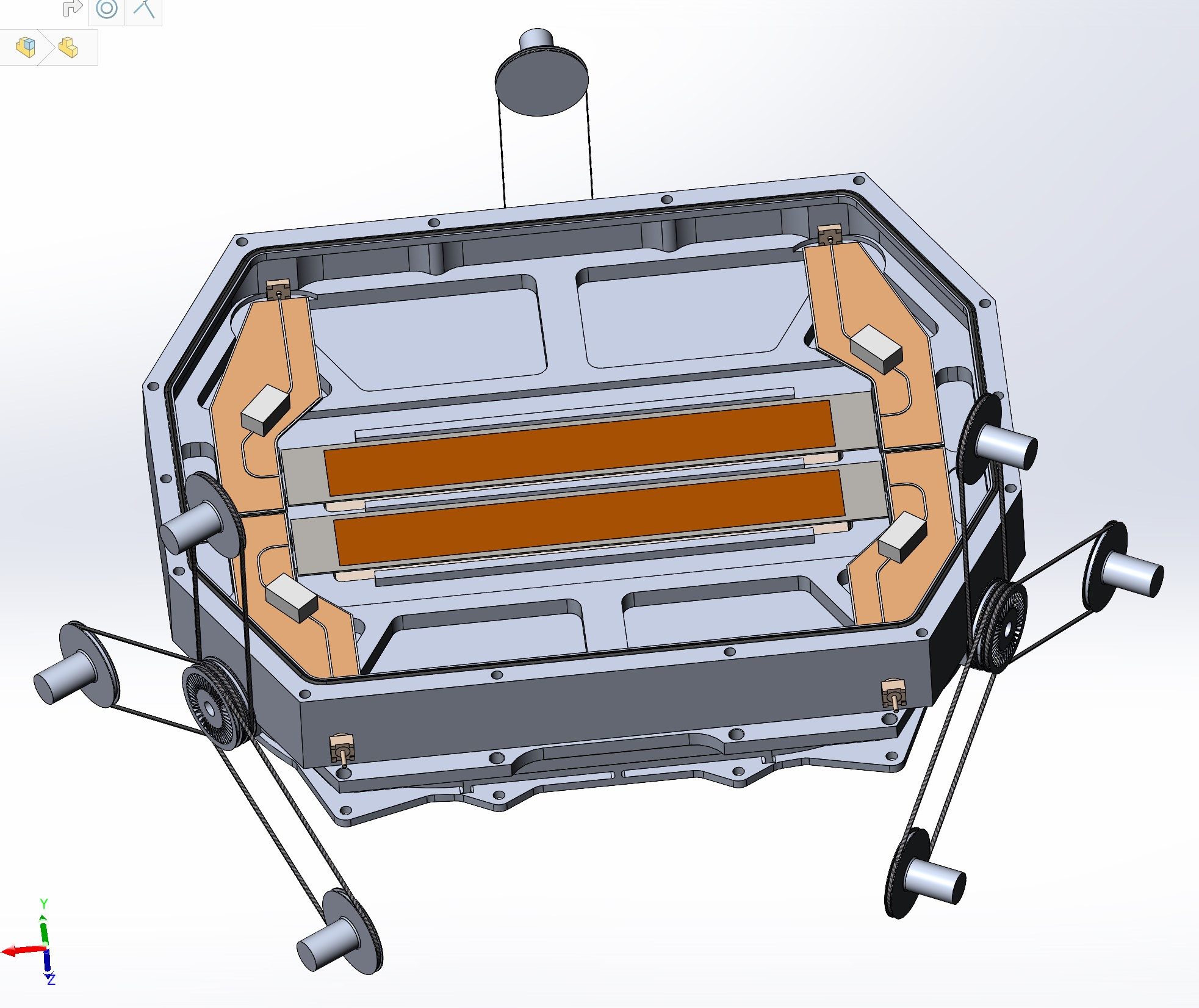}
\includegraphics*[clip, height=6.5cm]{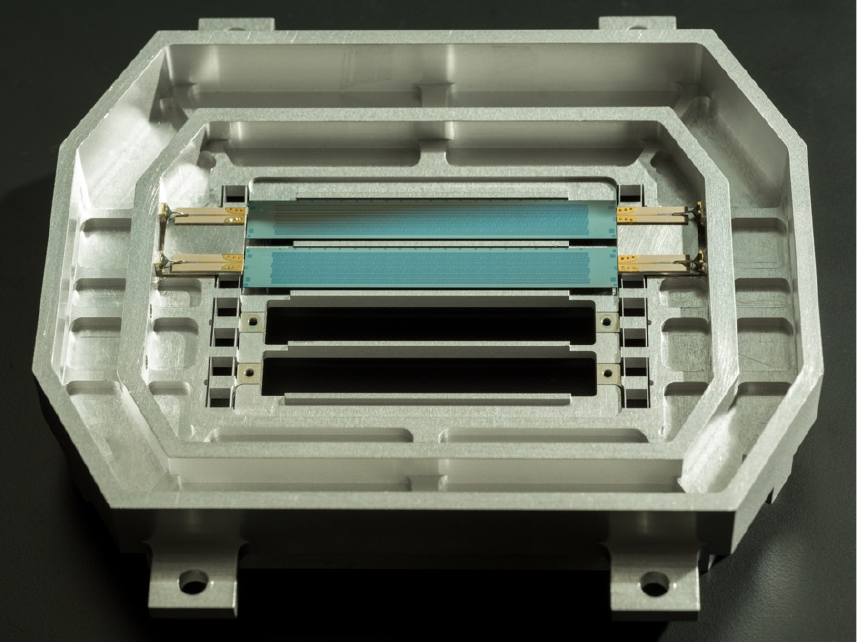}\\
\includegraphics*[clip, trim= 60 0 0 0, height=5.5cm]{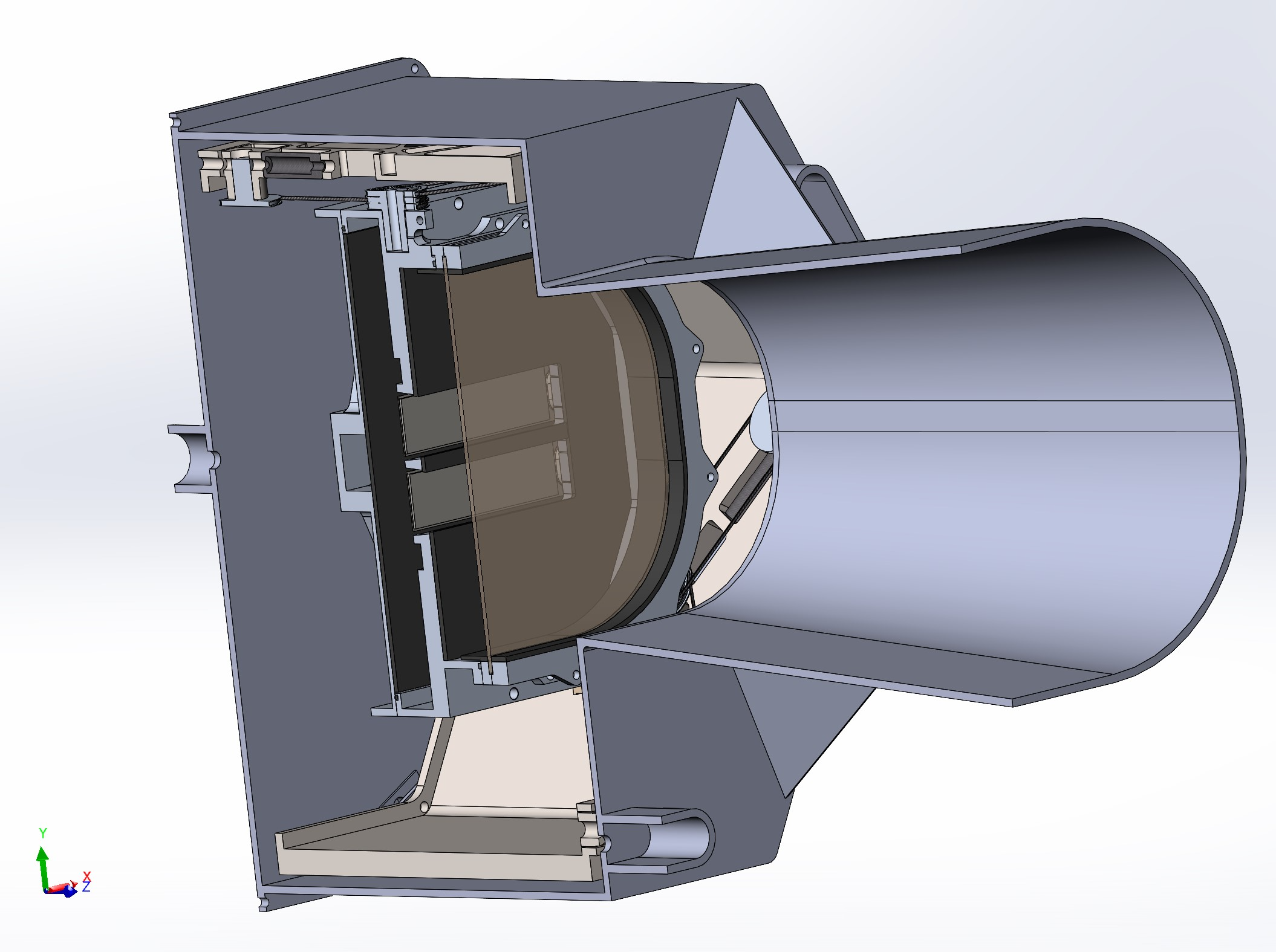}
\includegraphics*[clip, trim=780 20 50 450, height=5.5cm]{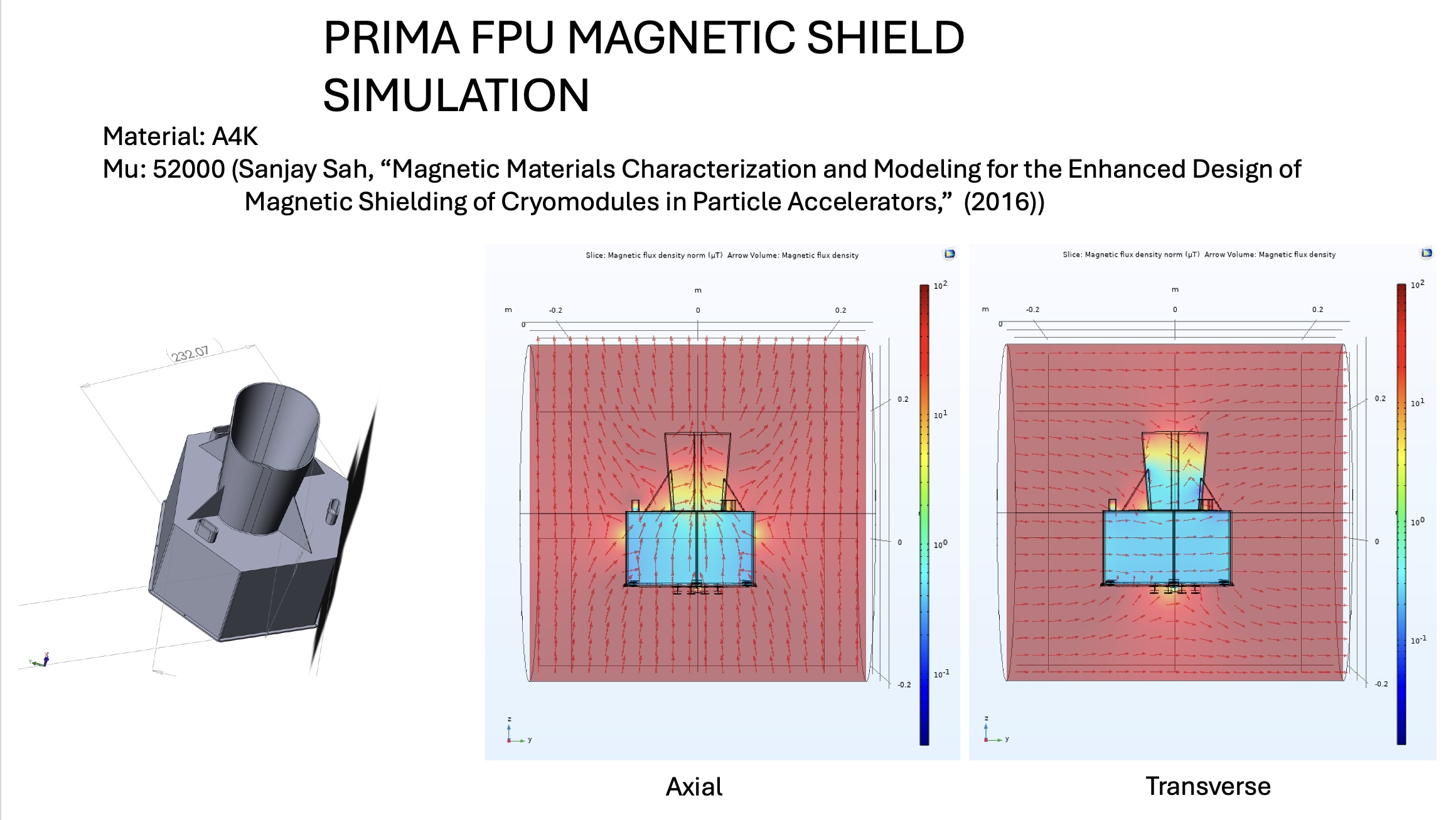}

    \caption[Array Suspension]{\label{fig:suspension}
FIRESS array housing and suspension.
Each focal plane consists of two 12 $\times$ 84-pixel subarrays with a gap.  Top, left shows the design which feature dual embedded aluminum enclosures which house the subarray chips and filters.  The chips are mounted on flexures which provide compliance against CTE mismatch yet have high enough resonant frequency to comfortably handle launch conditions. Top, right shows a flight-like housing (built for environmental testing of subarrays) with two subarrays mounted, detector side up.  Lenses are on the opposite side.    Bottom shows a cutaway view of the package including the 1~K side which both mounts the Kevlar suspension and the high-$\mu$ magnetic shield.  Right shows magnetic shielding simulations indicating shielding factors of 200 in both dimensions, more than is required.
}
    \end{center}
\end{figure}

The thermal suspension being designed for the FIRESS focal planes is shown in Figure~\ref{fig:suspension}.  It builds on JPL's development of the Herschel SPIRE bolometer suspension \cite{Hargrave_06,Griffin_10}, with tensioned Kevlar strings.  However, FIRESS uses a more modular approach, with nine individual bands, one for each force direction, rather than a single band looped through many pulleys and capstans.  The housing at 1 K provides the pre-load capability and compensates for the small negative coefficient of thermal expansion (CTE) of the Kevlar.  Unlike Herschel, PRIMA launches warm, so the challenging mechanical requirements must only be met at room temperature.  The design shown in Figure~\ref{fig:suspension} simultaneously meets the expected launch mechanical requirements and the thermal isolation requirements with margin.

\paragraph{Magnetic Shielding} By comparison with transition-edge sensor(TES) devices, KIDs have low sensitivity to magnetic fields. KIDs in our test facilities typically operate with modest shielding factors of $\sim$30--50 of Earth's field. This is sufficient to enable quality factors in aluminum above 10$^5$, ample for PRIMA.  As Figure~\ref{fig:suspension} shows, our design readily meets this requirement.

Conservative signal contamination considerations also create a 30--50$\times$ shielding requirement.  Small-signal AC fields \textcolor{black}{in PRIMA's signal frequency band (Table~\ref{tab:mod})} which could mimic astrophysical signals have been considered from various sources: the facility cooler, the beam steering mirror, the \textcolor{black}{continuous adiabatic demagnetization refrigerator (CADR)}, and the other elements of the spacecraft bus.  We must ensure that any such small signals are negligible even in long integrations; we adopt 12 hours, for which the RMS fractional frequency signal of the KIDs can integrates down to $2.6\times10^{-11}\,\mathrm{Hz/Hz}$.  An upper bound to the response to magnetic fields is 0.03~Hz/Hz per Tesla, based on measurements at JPL and SRON.  The largest exported field at the arrays is expected to be from the cooler compressor motor, for which an upper bound without shielding is 0.03~$\mu$T.  Full modulation of this field would produce a fractional frequency signal of $9\times10^{-10}\,\mathrm{Hz/Hz}$, some 35$\times$ higher than our imposed limit.  Shielding at this level would eliminate this concern.  We emphasize that this estimate is very conservative since 1) any such coherent magnetic signal will have a distinct signature in audio frequency space, so can be identified and removed, 2) Correlation analyses across all four arrays will likely reveal magnetic disturbances since they should create a spatial-spectra signature distinct from the astrophysical sources, and finally 3) we baseline \textcolor{black}{a niobium shield around the} focal plane package \textcolor{black}{, extending out from the focal plane, inside the high-mu shield. This shielding is not included in the simulations in Figure~\ref{fig:suspension}; it will} greatly ease the AC exported field requirement.

\begin{figure}[t!]
    \begin{center}
\includegraphics*[clip, trim=70 10 40 30,width=8.5cm]{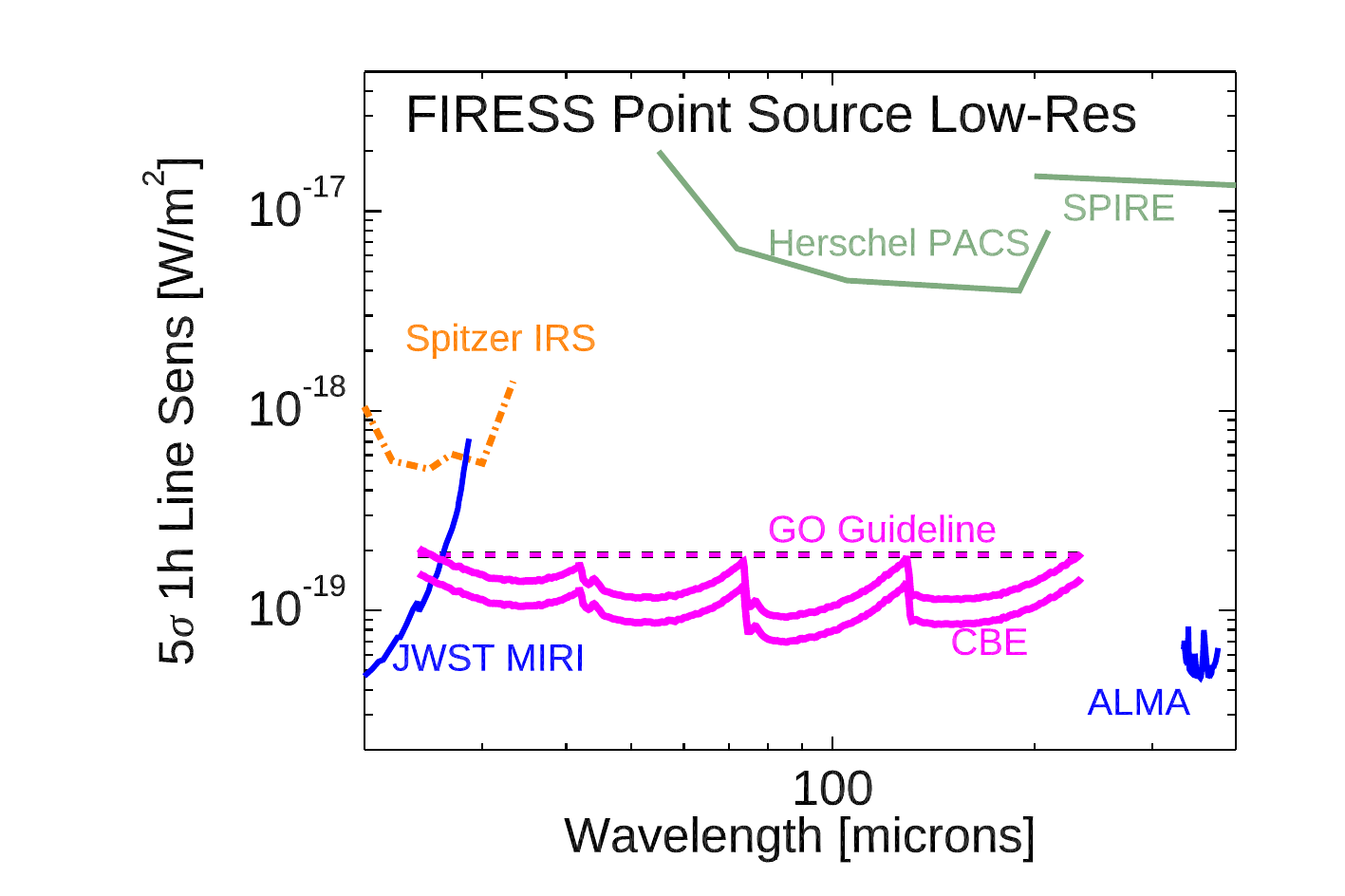}
\includegraphics*[clip, trim= 70 10 40 30, width=8.5cm]{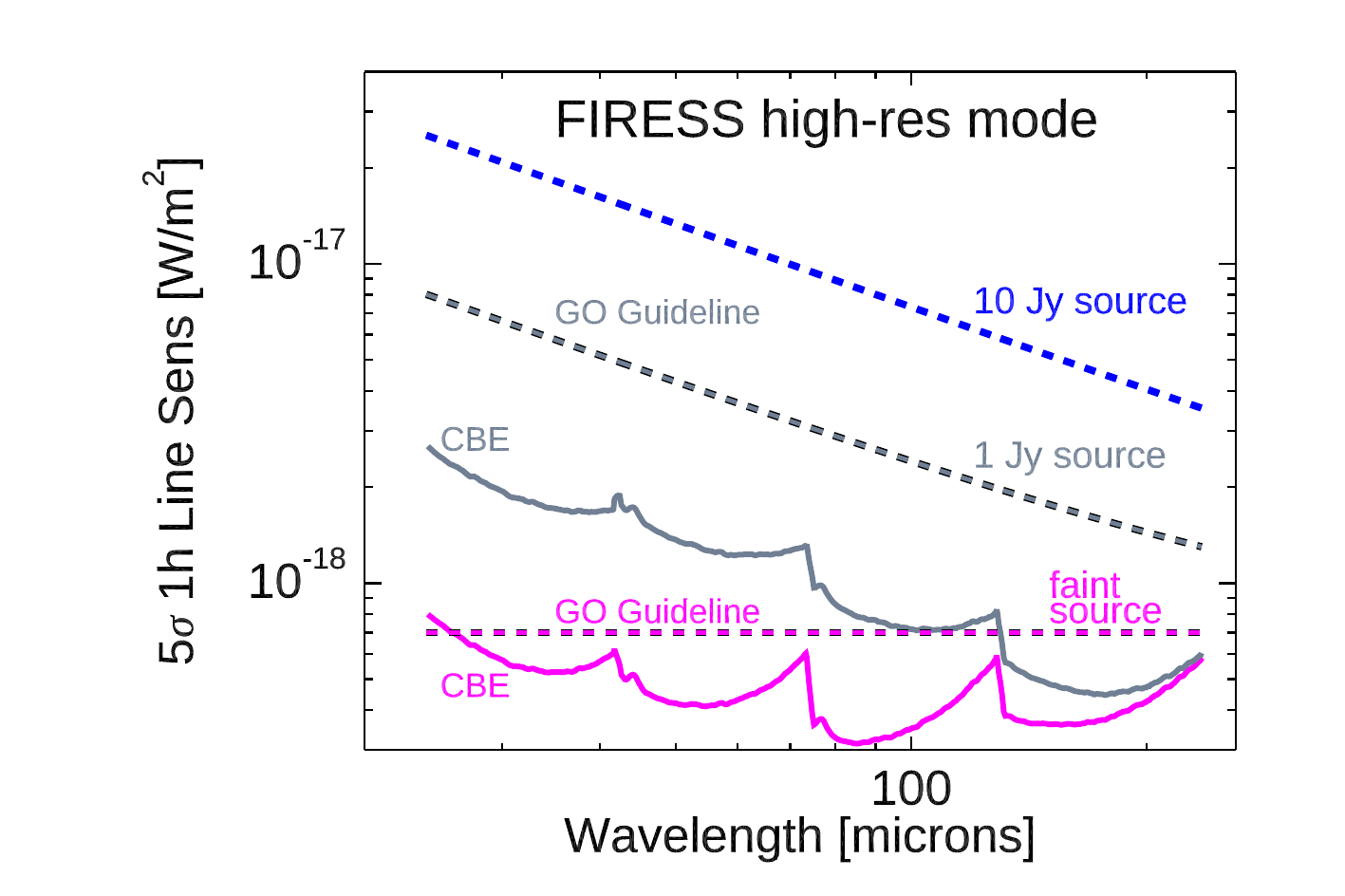}

    \caption[Performance Model]{\label{fig:pointed}PRIMA FIRESS sensitivity for pointed observations.  The \textcolor{black}{dashed}  curves \textcolor{black}{are the recommended performance to use for GO program development}. This is approximately the union of the requirements for the PI objectives, but extended across the full FIRESS band. The  \textcolor{black}{solid} magenta curve shows our current best estimate (CBE) of the performance, described in the text. The wavelength dependence in this estimate derives from the pixel coupling efficiency, especially important for pointed measurements.  \textcolor{black}{At left, we show two variations of the CBE which carry different assumptions about instrumental efficiencies; all subsequent plots use the most conservative (upper) estimates.}  At right, the estimate for the high-res mode with the FTM includes the impact of source photon noise, here the flux fiducial is applied at each wavelength.  The CBE is shown for the faint-source limit, and for the 1 Jy source, for which some water features from our disk model are visible in the noise at short wavelengths.}
    \end{center}
\end{figure}

\section{Performance Model} \label{sec:performance}
\begin{figure}[t!]
    \begin{center}
\includegraphics*[clip, trim=20 10 30 10,width=8cm]{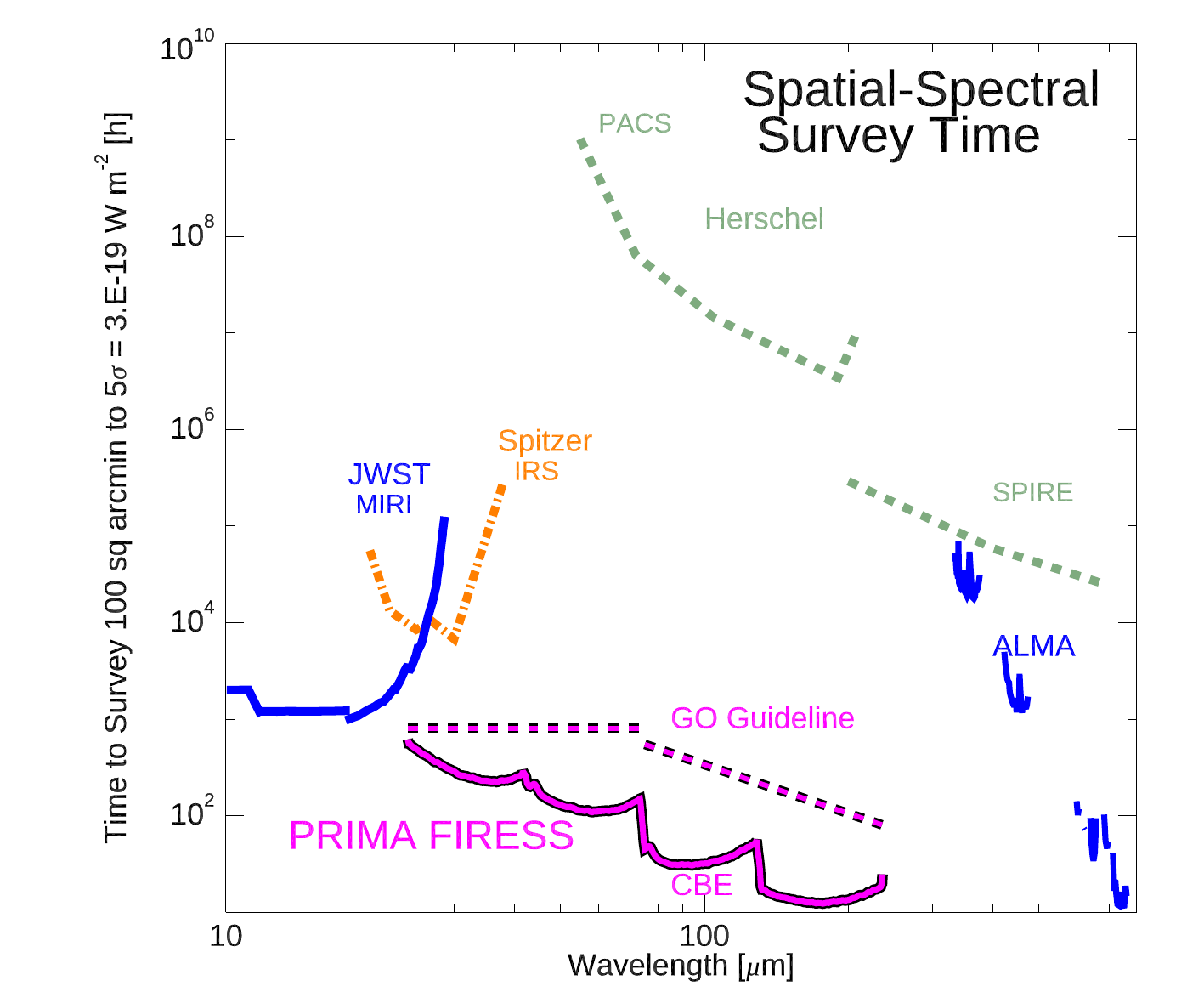}
\caption[Mapping Speed]{\label{fig:mapping}
    FIRESS mapping speed. Plotted is the time required to reach a depth of 3$\times 10^{-19}\,\rm W\,m^{-2}$ in unresolved line emission over a 100-square-arcsecond patch of sky \textcolor{black}{and with full-band coverage}.  The \textcolor{black}{dashed} curve for FIRESS tracks to the pointed-mode guideline sensitivity shown in Figure~\ref{fig:pointed}, and includes a provision for overheads in mapping. We recommend using this for scoping GO programs.  The \textcolor{black}{solid} magenta curve shows the current best estimate (CBE) \textcolor{black}{which carries some overheads described in the text.}  For FIRESS, time can be scaled as area / depth$^2$ for any desired program, \textcolor{black}{but with a suggested minimum area of 8 square arcminutes (see Section~\ref{sec:cbe}).}  Note that JWST MIRI time is dominated by overheads to cover the 100 square arcminutes in this fiducial; MIRI is much more favorable for smaller regions and/or deeper measurements than this comparison would suggest.}
    \end{center}    
\end{figure}

We have developed a performance model for FIRESS based on current best estimates \textcolor{black}{(CBE) of the key parameters.  This CBE is shown in Figures~\ref{fig:pointed} and \ref{fig:mapping} as the solid magenta curves with structure.  Also plotted in dashed curves and encapsulated in equations below are what we call the `GO guidelines.'  These are easy-to-use fiducials that we recommend for use in developing GO programs.  This performance meets or exceeds that required for the ensemble of PI science objectives at their various wavelengths, with simple interpolation through the FIRESS bands.} 

\textcolor{black}{\subsection{GO Guideline Sensitivities}}
We recommend using the `GO guidelines' for scoping potential general-observer programs, as the CBE is subject to evolution as the design matures.  All sensitivities refer to unresolved (narrow) spectral lines \textcolor{black}{in units of $\rm W\, m^{-2}$}.  Figure~\ref{fig:pointed} refers to point-source spectroscopy.  The left panel is for low-res mode, for which the BSM is chopping the source back and forth along the slit.  In this mode the \textcolor{black}{key metric is the the minimum detectable line flux (MDLF),} in an integration time t.  \textcolor{black}{It is a single number across the full FIRESS band:}
\begin{equation}
  \rm MDLF\,(5\sigma) \,\left[W\,m^{-2}\right] = 1.9\times10^{-19}\,\times \, \sqrt{\frac{1\, hour}{t}} .
\end{equation}

\textcolor{black}{The right panel of Figure~\ref{fig:pointed} shows the performance in the high-res mode. We anticipate that this mode will often be used with bright sources, so } we include the effect of the photon noise from the source, and express the required performance in an integration time t as a quadrature sum of the faint source limit and the source contribution:

\begin{equation}
  \rm  MDLF \, (5\sigma) \, \left[W\,m^{-2}\right] = \left( (7\times10^{-19})^2 + \left(8\times10^{-18} \times \sqrt{S} \times  \left( \frac{\lambda}{24~\mu m} \right)^{-0.87} \right)^2 \right)^{0.5}  \times \sqrt{\frac{1\, hour}{t}}.
\end{equation}
where $\rm S$ is the source flux in Janskys \textcolor{black}{at the wavelength of interest ($\lambda$)}, and the 2nd term accounts for the photon noise from the source.  This is plotted is shown in right panel of  Figure~\ref{fig:pointed}, for values of $\rm S$ \textcolor{black}{of 1~Jy and 10~Jy}.  A point to bear in mind for pointed measurements (both low-res and high-res) is this: while all four arrays are read out at once, the slits overlap in pairs (Section~\ref{sec:gratings}).  Thus, for a point source of interest, two of the four bands are on-source simultaneously, and obtaining full band spectra require $\sim$double the time.  

Figure~\ref{fig:mapping} refers to low-res spectral mapping; it shows the time required to reach a depth (5$\sigma$) of $3\times10^{-19}\,\rm W\,m^{-2}$ for unresolved spectral lines over a 100-square-arcminute area.  The expression capturing the  \textcolor{black}{GO guideline} expressed as a time for this measurement is:
\begin{equation}
\begin{split}
\rm t_{fidcial}\, [hours] = \ \ &800 \ \ \ \ \ \ \ \ \ \ \ \ \ \ \ \ \ \ \ \ \ \ \ \ \ \ \ \ \ \ \ \ 24 \mu m \le \lambda < 75 \mu m \\
= \ \ &336 \times \left(\frac{\lambda}{100 \mu m}\right)^{-1.68}  \ \ \ \ \ \ \ 75 \mu m \le \lambda \le 235 \mu m
\end{split}
\end{equation}
To estimate the time required for a different area and/or depth, apply the usual scalings: 
\begin{equation}
\rm time = t_{fiducial} \times \left(\frac{3\times10^{-19}\, W\,m^{-2}}{depth\, [5\sigma]}\right)^2 \,\times \left(\frac{Area}{100\,sq\,arcmin}\right)
\end{equation}
 The maps are obtained by rastering the slits around the sky using the observatory scan and/or the beam steering mirror (BSM) \textcolor{black}{(see Section~\ref{sec:modes}).  Given the sizes of the slits and their separation, we recommend using an area of no less than 8 square arcminutes in the calculator.   Smaller maps will need to be designed individually with band-by-band coverage taken into account. For large maps, the 3.3-arcminute slit separation is negligible.  }


\textcolor{black}{\subsection{Current Best Estimates}\label{sec:cbe}}
The CBE sensitivities shown in Figures~\ref{fig:pointed}~and~\ref{fig:mapping} are based on a detector intrinsic noise equivalent power (NEP) of \textcolor{black}{$0.9\times10^{-19}\,\rm W\,Hz^{-1/2}$}.  The model includes the photon noise from the scene and source (where applicable). For the faint source limit, this is based on an annual-average sightline toward the north ecliptic pole as viewed from \textcolor{black}{the sun-earth L2 point.} The optical efficiency is important; we use a total transmission from the sky to the focal plane of 11--15\%, with the variation due to the wavelength dependence of the grating blaze efficiency.  This includes the lens-to-absorber coupling of the KID array.  An additional factor accounts for the \textcolor{black}{diffractive losses, essentially the} coupling of a narrow-band spectral feature in a point source coupling through the slit and to a single pixel's 900-micron lens aperture.  This `pixel coupling factor' ranges from $\sim$ 0.15 to $\sim$ 0.58, with the larger numbers only found in Band 1.   Its variation accounts for \textcolor{black}{the sensitivity worsening working toward the long-wavelength ends of each band (pixel coupling drops toward long wavelengths).} \textcolor{black}{The pixel coupling factor is based on diffraction model for the instrument, it represents an average over wavelengths at each pixel, meaning it does not require the line to be centered with respect to the spectral position of a pixel.  Additionally, it includes the small (few percent at most) benefit of nearby spatial pixels contributing to the signal and being optimally combined, for \textcolor{black}{further} discussion, see the article by van Berkel et al. in this special issue.  For the purposes of sensitivity, our pixel coupling factor is equivalent to 1/$\sqrt{\rm N_{\rm eff}}$, where N$_{\rm eff}$ is the effective number of noise pixels used in source extraction.}

\textcolor{black}{The CBE sensitivities are based on a Monte-Carlo simulation of a mock observation of spectral lines centered at a large number of wavelengths across the band.  The simulation uses both types of spectral rows described in Section~\ref{sec:gratings}, each for half the integration time.}   Finally, the estimate includes temporal efficiencies of 88\% to account for the BSM chopping duty cycle and 90\% for lost data due to cosmic ray events (a conservative figure; see Kane et al.~2024 \cite{Kane_24}).  

For the high-res mode, the CBE includes the transmission penalties incurred in directing the light through the FTM,  both due to losses in the optics, as well as the 2$\times$ for the encoding with the FTS as the signal modulates from zero to full as the OPD is scanned.  The CBE shown here is conservative in that it does not include the benefit of the reduced photon noise on the detectors due to the reduced optical load.  This effect is included in the simulation presented in Section~\ref{sec:ftm}, and explains why the simulation outperforms this CBE estimate by a small factor.

\textcolor{black}{For the low-res mapping time, the CBE time is conservative.  It begins with the worst-case CBE sensitivities for pointed low-res mode, and assumes an instantaneous field of view (FoV) given by $24 \times \Omega_{\rm pix}$, where $\Omega_{\rm pix} = \theta_{\rm FWHM}^2\,\frac{\pi}{4 \ln{2}}$ is the (wavelength dependent) solid angle of given spatial pixel.
This estimate does not include the benefit of the slit being wider than a single pixel, which will increase the effective FoV by 1.3--1.6.  Finally, a factor of 3.2 is multiplied into the time, to account generously for mapping overheads and the impact of imperfect pixel yield in the arrays.  We expect to refine these estimates for the better as our design matures.}

\section{Summary}
The Far-Infrared Enhanced Survey Spectrometer (FIRESS) for PRIMA offers excellent capability with a manageable level of complexity.  Four slit-fed grating modules provide and combine to span the full 24-235~\mm\ range at R$\sim$100.  Each band has a 2016-pixel array of kinetic inductance detectors with noise subdominant to the natural photon background of the zodiacal light.   The result is a combination of full band coverage and spatial multiplexing with good sensitivity for both pointed measurements and spectroscopic mapping.  To \textcolor{black}{enhance} the resolving power to R$>$2000 across the band, a Fourier transform module is employed over a portion of the slits. The FTM uses the grating modules as the detector, an approach that offers full-band capability while preserving good line sensitivity.  This work presents these basic design elements, a simulation of the FIRESS-FTM + grating measurement, performance estimates, and a look at the key system-level requirements imposed by the detector system which we are addressing in our Phase-A design activity.

\newpage

\subsection* {Disclosures}

Copyright 2025. All rights reserved.

The authors declare that there are no financial interests, commercial affiliations, or other potential conflicts of interest that could have influenced the objectivity of this research or the writing of this paper.

\subsection* {Code and Data Availability}
The data that support the findings of this article are not publicly available due to the competition sensitive nature of the project at this time. They can be requested from the author at \tt{bradford@caltech.edu}.


\subsection* {Acknowledgments}
The research was carried out at the Jet Propulsion Laboratory, California Institute of Technology, under a contract with the National Aeronautics and Space Administration (80NM0018D0004)


\bibliography{report,bradford_master_bibdesk}   
\bibliographystyle{spiejour}   



\vspace{1ex}
\noindent Biographies and photographs of the other authors are not available.

\listoffigures
\listoftables

\end{document}

%% file: globalcommands.tex
\usepackage{amsmath}
\usepackage{amssymb}

\newcommand{\ms}{M$_{\odot}$}

\newcommand{\mm}{$\,\rm\mu m$}

\newcommand{\hh}{H$_2$}

\newcommand{\be}{\begin{equation}}
\newcommand{\ee}{\end{equation}}

\newcommand{\too}{\!\rightarrow\!} 
\newcommand{\jone}{\small{$\rm J\eqq1\too0$}\normalsize}

\newcommand{\Res}{\(\mathcal{R}\)}

\newcommand{\by}{$\times$}

\newcommand{\eqq}{\!=\!} 

%% file: article.bbl
\begin{thebibliography}{10}

\bibitem{MadauDickinson_14}
P.~{Madau} and M.~{Dickinson}, ``{Cosmic Star-Formation History},'' {\em \araa} {\bf 52}, 415--486  (2014).

\bibitem{Watson_2015_earlydust}
D.~{Watson}, L.~{Christensen}, K.~K. {Knudsen}, {\em et~al.}, ``{A dusty, normal galaxy in the epoch of reionization},'' {\em \nat} {\bf 519}, 327--330  (2015).

\bibitem{Tamura_2019_oiii_dust}
Y.~{Tamura}, K.~{Mawatari}, T.~{Hashimoto}, {\em et~al.}, ``{Detection of the Far-infrared [O III] and Dust Emission in a Galaxy at Redshift 8.312: Early Metal Enrichment in the Heart of the Reionization Era},'' {\em \apj} {\bf 874}, 27  (2019).

\bibitem{Witstok:2023_carbon_dust}
J.~Witstok, I.~Shivaei, R.~Smit, {\em et~al.}, ``Carbonaceous dust grains seen in the first billion years of cosmic time,'' {\em Nature} {\bf 621}(7978), 267--270  (2023).

\bibitem{Zavala_2021_IRLumFun}
J.~A. {Zavala}, C.~M. {Casey}, S.~M. {Manning}, {\em et~al.}, ``{The Evolution of the IR Luminosity Function and Dust-obscured Star Formation over the Past 13 Billion Years},'' {\em \apj} {\bf 909}, 165  (2021).

\bibitem{Hickox_Alexander_2018}
R.~C. {Hickox} and D.~M. {Alexander}, ``{Obscured Active Galactic Nuclei},'' {\em \araa} {\bf 56}, 625--671  (2018).

\bibitem{Stacey_93}
G.~J. {Stacey}, ``{Far-Infrared Spectroscopy Diagnostics of the Interstellar Medium in Galaxies},'' in {\em Astronomical Infrared Spectroscopy: Future Observational Directions},  S.~{Kwok}, Ed., {\em Astronomical Society of the Pacific Conference Series} {\bf 41}, 297  (1993).

\bibitem{Sturm_2002}
E.~{Sturm}, D.~{Lutz}, A.~{Verma}, {\em et~al.}, ``{Mid-Infrared line diagnostics of active galaxies. A spectroscopic AGN survey with ISO-SWS},'' {\em \aap} {\bf 393}, 821--841  (2002).

\bibitem{Stone_2022}
M.~{Stone}, A.~{Pope}, J.~{McKinney}, {\em et~al.}, ``{Measuring Star Formation and Black Hole Accretion Rates in Tandem Using Mid-infrared Spectra of Local Infrared Luminous Galaxies},'' {\em \apj} {\bf 934}, 27  (2022).

\bibitem{Bergin_13}
E.~A. {Bergin}, L.~I. {Cleeves}, U.~{Gorti}, {\em et~al.}, ``{An old disk still capable of forming a planetary system},'' {\em \nat} {\bf 493}, 644--646  (2013).

\bibitem{Seo_2024_disksHD}
Y.~M. {Seo}, K.~{Willacy}, G.~{Bryden}, {\em et~al.}, ``{Retrievals of Protoplanetary Disk Parameters Using Thermochemical Models. I. Disk Gas Mass from Hydrogen Deuteride Spectroscopy},'' {\em \apj} {\bf 967}, 131  (2024).

\bibitem{Banzatti_2025_JWSTdiskWater}
A.~{Banzatti}, C.~{Salyk}, K.~M. {Pontoppidan}, {\em et~al.}, ``{Water in Protoplanetary Disks with JWST-MIRI: Spectral Excitation Atlas and Radial Distribution from Temperature Diagnostic Diagrams and Doppler Mapping},'' {\em \aj} {\bf 169}, 165  (2025).

\bibitem{Nakagawa_02}
T.~{Nakagawa} and {Spica Working Group}, ``{SPICA: a new space telescope for mid- and far-infrared astronomy},'' {\em Advances in Space Research} {\bf 30}, 2129--2134  (2002).

\bibitem{Bradford_06}
C.~M. {Bradford}, ``{BLISS for SPICA: far-IR spectroscopy at the background limit},'' in {\em Society of Photo-Optical Instrumentation Engineers (SPIE) Conference Series},  {\em Society of Photo-Optical Instrumentation Engineers (SPIE) Conference Series} {\bf 6265}, 0  (2006).

\bibitem{Roelfsema_18}
P.~R. {Roelfsema}, H.~{Shibai}, L.~{Armus}, {\em et~al.}, ``{SPICA-A Large Cryogenic Infrared Space Telescope: Unveiling the Obscured Universe},'' {\em \pasa} {\bf 35}, e030  (2018).

\bibitem{Meixner_19_Origins}
M.~{Meixner}, A.~{Cooray}, D.~{Leisawitz}, {\em et~al.}, ``{Origins Space Telescope Mission Concept Study Report},'' {\em arXiv e-prints} , arXiv:1912.06213  (2019).

\bibitem{Bradford_2021_OSS}
C.~M. Bradford, B.~A. Cameron, B.~D. Moore, {\em et~al.}, ``{Origins Survey Spectrometer: revealing the hearts of distant galaxies and forming planetary systems with far-IR spectroscopy},'' {\em Journal of Astronomical Telescopes, Instruments, and Systems} {\bf 7}(1), 1 -- 28  (2021).

\bibitem{Jellema_17}
W.~{Jellema}, C.~{Pastor}, D.~{Naylor}, {\em et~al.}, ``{Safari: instrument design of the far-infrared imaging spectrometer for spica},'' in {\em Society of Photo-Optical Instrumentation Engineers (SPIE) Conference Series},  {\em Society of Photo-Optical Instrumentation Engineers (SPIE) Conference Series} {\bf 10563}, 105631K  (2017).

\bibitem{HaileyDunsheath_18}
S.~{Hailey-Dunsheath}, A.~C.~M. {Barlis}, J.~E. {Aguirre}, {\em et~al.}, ``{Development of Aluminum LEKIDs for Balloon-Borne Far-IR Spectroscopy},'' {\em Journal of Low Temperature Physics}   (2018).

\bibitem{Janssen2021JLTP}
R.~M.~J. {Janssen}, R.~{Nie}, B.~{Bumble}, {\em et~al.}, ``{Single pixel performance of the kinetic inductance detectors for the Terahertz Intensity Mapper},'' {\em Journal of Low Temperature Physics} , in review  (2021).

\bibitem{Hailey_2023}
S.~{Hailey-Dunsheath}, S.~{van Berkel}, A.~E. {Beyer}, {\em et~al.}, ``{Characterization of a Far-Infrared Kinetic Inductance Detector Prototype for PRIMA},'' {\em arXiv e-prints} , arXiv:2311.03586  (2023).

\bibitem{Day_2024}
P.~K. {Day}, N.~F. {Cothard}, C.~{Albert}, {\em et~al.}, ``{A 25-micron single photon sensitive kinetic inductance detector},'' {\em arXiv e-prints} , arXiv:2404.10246  (2024).

\bibitem{Foote_24}
L.~{Foote}, C.~{Albert}, J.~{Baselmans}, {\em et~al.}, ``{High-Sensitivity Kinetic Inductance Detector Arrays for the PRobe Far-Infrared Mission for Astrophysics},'' {\em Journal of Low Temperature Physics} {\bf 214}, 219--229  (2024).

\bibitem{Kane_24}
E.~{Kane}, C.~{Albert}, J.~{Baselmans}, {\em et~al.}, ``{Modeling of Cosmic Rays and Near-IR Photons in Aluminum KIDs},'' {\em Journal of Low Temperature Physics} {\bf 214}, 238--246  (2024).

\bibitem{Krause_06}
O.~{Krause}, D.~{Lemke}, R.~{Hofferbert}, {\em et~al.}, ``{The cold focal plane chopper of HERSCHEL's PACS instrument},'' in {\em Optomechanical Technologies for Astronomy},  E.~{Atad-Ettedgui}, J.~{Antebi}, and D.~{Lemke}, Eds., {\em Society of Photo-Optical Instrumentation Engineers (SPIE) Conference Series} {\bf 6273}, 627325  (2006).

\bibitem{Houck_04}
J.~R. {Houck}, T.~L. {Roellig}, J.~{Van Cleve}, {\em et~al.}, ``{The infrared spectrograph on the Spitzer Space Telescope},'' in {\em Optical, Infrared, and Millimeter Space Telescopes},  J.~C. {Mather}, Ed., {\em \procspie} {\bf 5487}, 62--76  (2004).

\bibitem{Bouchet_2015}
P.~{Bouchet}, M.~{Garc{\'\i}a-Mar{\'\i}n}, P.~O. {Lagage}, {\em et~al.}, ``{The mid-infrared instrument for the James Webb Space Telescope, III: MIRIM, the MIRI imager}.'' Technical Report JWST-STScI-000003  (2015).

\bibitem{Fixsen_94}
D.~J. {Fixsen}, E.~S. {Cheng}, D.~A. {Cottingham}, {\em et~al.}, ``{Calibration of the COBE FIRAS instrument},'' {\em \apj} {\bf 420}, 457--473  (1994).

\bibitem{Griffin_10}
M.~J. {Griffin}, A.~{Abergel}, A.~{Abreu}, {\em et~al.}, ``{The Herschel-SPIRE instrument and its in-flight performance},'' {\em \aap} {\bf 518}, L3+  (2010).

\bibitem{Nikola_VIPA_24}
T.~{Nikola}, B.~{Zou}, G.~J. {Stacey}, {\em et~al.}, ``{Virtually Image Phased Array (VIPA): demonstration of the next generation direct detection spectrometer for velocity resolved spectroscopy in the far-infrared},'' in {\em Millimeter, Submillimeter, and Far-Infrared Detectors and Instrumentation for Astronomy XII},  J.~{Zmuidzinas} and J.-R. {Gao}, Eds., {\em Society of Photo-Optical Instrumentation Engineers (SPIE) Conference Series} {\bf 13102}, 131020F  (2024).

\bibitem{Martin_70}
D.~Martin and E.~Puplett, ``Polarised interferometric spectrometry for the millimetre and submillimetre spectrum,'' {\em Infrared Physics} {\bf 10}(2), 105 -- 109  (1970).

\bibitem{Lambert_78}
D.~K. Lambert and P.~L. Richards, ``Martin-puplett interferometer: an analysis,'' {\em Appl. Opt.} {\bf 17}, 1595--1602  (1978).

\bibitem{mather_93}
J.~C. {Mather}, D.~J. {Fixsen}, and R.~A. {Shafer}, ``{Design for the COBE far-infrared absolute spectrophotometer (FIRAS)},'' in {\em Infrared Spaceborne Remote Sensing},  M.~S. {Scholl}, Ed., {\em Society of Photo-Optical Instrumentation Engineers (SPIE) Conference Series} {\bf 2019}, 168--179  (1993).

\bibitem{Cournoyer_22}
A.~{Cournoyer}, {\'E}.~{Carbonneau}, P.~{Gilbert}, {\em et~al.}, ``{Cryogenic testing towards TRL-5 demonstration of a novel stiffness-compensated, reactionless scan mechanism for the Fourier transform spectrometer of SPICA SAFARI instrument},'' in {\em Millimeter, Submillimeter, and Far-Infrared Detectors and Instrumentation for Astronomy XI},  J.~{Zmuidzinas} and J.-R. {Gao}, Eds., {\em Society of Photo-Optical Instrumentation Engineers (SPIE) Conference Series} {\bf 12190}, 121902E  (2022).

\bibitem{Cournoyer_20}
A.~{Cournoyer}, {\'E}.~{Carbonneau}, P.~{Gilbert}, {\em et~al.}, ``{Design of a novel cryogenic stiffness-compensated reactionless scan mechanism for the Fourier transform spectrometer of SPICA SAFARI instrument},'' in {\em Millimeter, Submillimeter, and Far-Infrared Detectors and Instrumentation for Astronomy X},  J.~{Zmuidzinas} and J.-R. {Gao}, Eds., {\em Society of Photo-Optical Instrumentation Engineers (SPIE) Conference Series} {\bf 11453}, 1145338  (2020).

\bibitem{Gratton_92}
R.~{Gratton} and R.~{Bhatia}, ``{Theory of the POST Dispersed Fourier Transform Spectrograph},'' in {\em European Southern Observatory Conference and Workshop Proceedings},  {\em European Southern Observatory Conference and Workshop Proceedings} {\bf 40}, 293  (1992).

\bibitem{Jellema_16}
W.~Jellema, D.~A. Naylor, and P.~Roelfsema, ``Post-dispersed fts spectroscopy on spica-safari,'' in {\em Light, Energy and the Environment},  {\em Light, Energy and the Environment} , FM2D.6, Optica Publishing Group  (2016).

\bibitem{Naylor_22}
D.~A. {Naylor}, B.~G. {Gom}, A.~M. {Anderson}, {\em et~al.}, ``{Development of a cryogenic far-infrared post-dispersed polarizing Fourier transform spectrometer},'' in {\em Millimeter, Submillimeter, and Far-Infrared Detectors and Instrumentation for Astronomy XI},  J.~{Zmuidzinas} and J.-R. {Gao}, Eds., {\em Society of Photo-Optical Instrumentation Engineers (SPIE) Conference Series} {\bf 12190}, 121900R  (2022).

\bibitem{Norton_76}
R.~H. Norton and R.~Beer, ``New apodizing functions for fourier spectrometry,'' {\em J. Opt. Soc. Am.} {\bf 66}, 259--264  (1976).

\bibitem{Cothard_24_lenses}
N.~F. {Cothard}, T.~{Stevenson}, J.~{Mateo}, {\em et~al.}, ``{Monolithic silicon microlens arrays for far-infrared astrophysics},'' {\em \ao} {\bf 63}, 1481  (2024).

\bibitem{Chen_23_PRIMAcryo}
W.~Chen, B.~Moore, M.~DiPirro, {\em et~al.}, ``{PRIMA space telescope cryocooling system},'' in {\em Infrared Sensors, Devices, and Applications XIII},  P.~Wijewarnasuriya, A.~I. D'Souza, and A.~K. Sood, Eds.,  {\bf 12687}, 1268705, International Society for Optics and Photonics, SPIE  (2023).

\bibitem{Hargrave_06}
P.~Hargrave, J.~Bock, C.~Brockley-Blatt, {\em et~al.}, ``{The 300-mK system for Herschel-SPIRE},'' in {\em Millimeter and Submillimeter Detectors and Instrumentation for Astronomy III},  J.~Zmuidzinas, W.~S. Holland, S.~Withington, {\em et~al.}, Eds.,  {\bf 6275}, 627513, International Society for Optics and Photonics, SPIE  (2006).

\end{thebibliography}
